\documentclass{aa}
\usepackage[varg]{txfonts}
\bibpunct{(}{)}{;}{a}{}{,} 
\usepackage{multirow}
\usepackage{graphicx}

\newcommand{\hii}{\ion{H}{ii}}
\newcommand{\hh}{\ion{H}{ii}~}
\newcommand{\hei}{\ion{He}{i}}

\newcommand{\nii}{[\ion{N}{ii}]}
\newcommand{\oi}{[\ion{O}{i}]}
\newcommand{\oii}{[\ion{O}{ii}]}
\newcommand{\oiii}{[\ion{O}{iii}]}
\newcommand{\sii}{[\ion{S}{ii}]}

\begin{document}

\title{Inner and outer star forming regions over the disks of spiral galaxies} 

\subtitle{I. Sample characterization} 

\author{M. Rodr\'{i}guez-Baras\inst{1} \and A.I. D\'{i}az\inst{1} \and F.F. Rosales-Ortega\inst{2} 
     \and S.F. S\'{a}nchez\inst{3}}

\institute{Departamento de F\'{i}sica Te\'orica, Universidad Aut\'onoma de Madrid, 28049 Madrid, Spain.
  \and Instituto Nacional de Astrof{\'i}sica, {\'O}ptica y Electr{\'o}nica, Luis E. Erro 1, 72840 Tonantzintla, Puebla, M\'exico.
  \and Instituto de Astronom\'{i}a, Universidad Nacional Aut\'onoma de M\'exico, A.P. 70-264, 04510 M\'exico D.F., M\'exico.} 

\date{\today}

\abstract {The knowledge of abundance distributions is central to
  understanding the formation and evolution of galaxies. Most of the relations
  employed for the derivation of gas abundances have so far been derived from
  observations of outer disk \hh regions, despite the known differences between
  inner and outer regions.} {Using integral field spectroscopy (IFS)
  observations we aim to perform a systematic study and comparison of two inner and
  outer \hh regions samples. The spatial resolution of the IFS, the number of
  objects and the homogeneity and coherence of the observations
  allow a complete characterization of the main observational properties and
  differences of the regions.} {We analyzed a sample of 725 inner \hh regions and
  a sample of 671 outer \hh regions, all of them detected and extracted from the
  observations of a sample of 263 nearby, isolated, spiral galaxies observed by
  the CALIFA survey. } {We find that inner \hh regions show smaller equivalent
  widths, greater extinction and luminosities, along with greater values
  of \nii~$\lambda$6583/H$\alpha$ and \oii~$\lambda$3727/\oiii~$\lambda$5007
  emission-line ratios, indicating higher metallicites and lower ionization
  parameters. Inner regions have also redder colors and higher photometric and
  ionizing masses, although M$_{ion}$/M$_{phot}$ is slighty higher for
  the outer regions.} {This work shows important observational differences
  between inner and outer \hh regions in star forming galaxies not previously
  studied in detail. These differences indicate that inner regions have more
  evolved stellar populations and are in a later evolution state with respect to
  outer regions, which goes in line with the inside-out galaxy formation
  paradigm.}

\maketitle

\section{Introduction}
\label{Introduction}

It is widely recognized that the knowledge of abundance distributions in galaxies is very important as a probe of their chemical evolution and star formation histories. \hh regions in external spiral and irregular galaxies provide an excellent means to derive the chemical abundances of different elements, both primordial and product of stellar nucleosynthesis. This information is central to guiding theoretical models of the formation and evolution of galaxies. 

Among the different abundance-related parameters employed are: (i) the radial metallicity gradient; (ii) the average metallicity at a given fiducial galactic radius; and (iii) the central metallicity value, where the term "metallicity" usually refers to oxygen abundance, oxygen being the most abundant element in the universe after hydrogen and helium. In fact, the two latter parameters rely on the determination of the first, since they are calculated either by interpolation or extrapolation of the metallicity distribution respectively. However, it should be kept in mind that what is referred to as "abundance" or "radial abundance gradient" of a galaxy in fact represents an observational limitation derived from the need to use fixed aperture and/or long-slit spectroscopy. The information actually required to deeply approach the issue of the formation and evolution of disk galaxies is the map of the abundance distribution of the different elements, that up to now represented a very difficult and highly time consuming task. New multi-slit and integral field spetroscopy data are now readily obtained, which have greatly increased the number of \hh regions analyzed per given galaxy, although our methodologies still suffer from essentially the same systematic uncertainties in the determination of gas abundance distributions. 

        In general terms, there are two different approaches to derive elemental gas abundances: the so called direct method, which makes use of the measurement of the electron temperature T$_e$ from the quotient of (faint) auroral to (strong) nebular lines of different elements (O, N, and S among others), and semi-empirical models, in which combinations of (strong) nebular lines are used to infer abundances in regions where the (faint) auroral lines cannot be detected through suitable calibrations. Due to the cooling properties of the nebulae, the regions to which the direct method is applicable are of low metallicity and therefore, in the case of spiral galaxies, tend to reside in the outer disc zones, while the regions that require the application of semi-empirical methods, being of higher metallicity, tend to reside in the most inner zones of the discs. These different treatments of outer and inner disc \hh regions can complicate the interpretation of the radial gas abundance gradients, and could even produce artificial effects if inner and outer \hh regions differ in physical properties and ionization structure.

        Although the emission line spectra of \hh regions in the inner and outer regions of disks look alike, some differences between these two families are recognized. Inner \hh regions seem to show lower \oiii~$\lambda$5007/H$\beta$ values than their outermost counterparts, an effect that can be produced by the combination of lower effective temperatures of their ionizing stars, higher dust content and higher metallicity; at the same time, \oii~$\lambda$3727/\oiii~$\lambda$5007 is higher in the spectra of the inner regions, which seems to indicate a lower excitation of the gas; \nii~$\lambda$6583/\oii~$\lambda$3727 is also higher, pointing to a larger N/O, also indicative of a higher overall metallicity \citep{1983ASIC..110..205P}. A hint of somewhat higher electron density of the emitting gas in the inner regions has also been reported \citep{1989AJ.....97.1022K,2004ApJ...615..228B,2007MNRAS.382..251D}. At high abundances, as those expected for inner or circumnuclear \hh regions, the density of the nebula affects significantly the strength of emission lines, specially \oiii, due to the competition between collisional and radiative de-excitation in the nebular cooling fine structure O$^{++}$ transitions \citep{1993ApJ...411..137O}. 

Higher extinctions, such as could be expected in higher metallicity and higher density regions, could also have an impact on the emission line intensities. The presence of dust can modify the thermal structure of nebulae in several ways. Firstly, the removal of cooling agents from the gas phase via depletion onto grains will increase the electron temperature \citep{1993MNRAS.261..306H}. Secondly, dust grains can absorb a given fraction of the Lyman continuum photons and thus modify the ionizing radiation field \citep{1986PASP...98..995M}. The absorbed energy will then be re-radiated in the IR \citep{1991A&AS...88..399M}. Heating or cooling by photo-emission or recombination from charged grains can also affect the thermal balance of the nebulae \citep{1991ApJ...374..580B}.

There is no doubt that the analysis of high excitation, low metallicity spectra is easier than that of the opposite case, and the larger contribution from the underlying stellar continuum in the case of the innermost regions represents a limitation. Therefore, despite any inferred different physical properties of inner and outer \hh regions, most of the relations currently employed for the derivation of abundances have been derived from observations of outer disk regions and have been assumed to be valid for all the ionized regions over the whole galactic disk. On the other hand, these inferences have been obtained from the study, albeit detailed, of a relatively small number of objects. 

Fortunately it is now possible with the advent of multi-object spectrographs (MOS) or integral field units (IFU) to perform a complete spectroscopic mapping of the distributions of \hh regions over the disks of spirals \citep[see e.g., ][]{2008ApJ...675.1213R,2011MNRAS.415.2439R,2015MNRAS.454.3664B}. A brief account of the results of this kind of work  includes the presence of a considerable dispersion in the derived abundances at a given galactocentric distance and the indication of possible azimuthal variations. The first could be due to the different sizes of the \hh regions observed, with the smallest regions being affected by stochasticity in the stellar mass function, and the second could be ascribed to differences in star formation in and between spiral arms and also to differences in mixing in the turbulent interstellar medium.

The recently completed CALIFA (Calar Alto Legacy Integral Field Area) survey provides an excellent opportunity to perform a systematic study of the properties of inner and outer \hh regions over the disks of spiral galaxies, since the homogeneity of the data regarding both observations and handling is a requirement to obtain reliable results. This in turn will give us the possibility of exploring the effects that any existing differences may have in derived properties of the regions themselves, such as elemental abundances, ionization structure, evolutionary state, amongst other. 

This is the first article of a series and presents the account of the observational properties of inner and outer regions in a sample of 263 nearby, isolated, spiral galaxies. In Section 2 we provide a summary of the observations on which the  work is based. Section 3 presents the characteristics of both the galaxy sample and the \hh region sample used. Section 4 presents the results of this characterisation, together with their discussion. Finally our conclusions are summarized in Section 5.

\begin{table*}
\caption{Physical properties of part of the CALIFA galaxies involved in this work, as described in the text. The complete table can be found in this paper online version. The corresponding sources are: (i) Galaxy name. (ii) Redshift. Given by the CALIFA survey, that obtained them from SIMBAD database on January 2010 (see W14). (iii) Morphological type. Own classification of the CALIFA survey, made by a combination of by-eye classification by five collaborators (see W14) (iv)
Inclination. From the axis ratios obtained by calculating light moments, and using the expression given by \cite{1958MeLuS.136....1H}. (v) Distance (Mpc). From distance moduli corrected for Virgo-centric infall from Hyperleda
catalog. (vi) Effective radius. Derived by the CALIFA survey, based on an analysis of the azimuthal surface brightness profile (more information in Sect. \ref{Effective radius}). (vii) Magnitudes g and r. From the Sloan Digital Sky Survey (SDSS). (viii) Absolute B magnitude. Calculated from the SDSS magnitudes, using Lupton (2005, see Sect. \ref{Inner regions sample}) conversion. (ix) Total number of \hh regions extracted in the galaxy.} 
\label{table_califa_galaxies}     
\centering                                     
\begin{tabular}{l c c c c c c c c c c c}          
\hline\hline                     
\noalign{\smallskip}

Galaxy & Redshift & Morph. type & Inclination & Distance & $\textit{R}_{eff}$ & g & r & $M_{B}$ & $N_{t}$\\
&&&&(Mpc)&(kpc)&&&&\\ 

\hline

\noalign{\smallskip}

NGC\,0001 & 0.015 & Sbc & 53.3 & 66.09 & 6.16 & 13.70 & 12.89 & -17.47 & 53\\ 

NGC\,0023 & 0.015 & Sb & 44.5 & 66.10 & 8.726 & 12.67 & 11.95 & -19.12 & 48\\ 

NGC\,0160 & 0.017 & Sa & 84.0 & 75.04 & 11.685 & 13.43 & 12.56 & -18.95 & 6\\

NGC\,0165 & 0.020 & Sb & 23.8 & 84.00 & 13.314 & 14.02 & 13.36 & -18.93 & 65\\ 

NGC\,0177 & 0.013 & Sab & 86.1 & 52.79 & 7.795 & 13.82 & 13.10 & -20.49 & 14\\ 

NGC\,0180 & 0.018 & Sb & 52.2 & 75.04 & 13.063 & 13.70 & 12.80 & -17.47 & 74\\ 

NGC\,0214 & 0.015 & Sbc & 69.8 & 66.08 & 6.933 & 12.86 & 12.17 & -18.27 & 42\\

NGC\,0216 & 0.005 & Sd & 79.2 & 21.86 & 2.049 & 13.67 & 13.21 & -18.52 & 22\\

\noalign{\smallskip}

\hline
\end{tabular}  
\end{table*}

\section{Summary of observations and data reduction}
\label{Observations}

The galaxies used in this work are part of the CALIFA project, one of the most ambitious 2D-spectroscopic surveys to date. The observations were carried out at the Centro Astron\'{o}mico Hispano-Alem\'{a}n (CAHA) 3.5 m telescope. This work is based on the 350 galaxies observed using the low-resolution setup until September 2014. Most of these galaxies are part of the 2nd CALIFA Data Release \citep[DR2, ][]{2015A&A...576A.135G}, and therefore the datacubes are accessible from the DR2 webpage\footnote{http://califa.caha.es/DR2}. The CALIFA survey is already finished, and the complete observations are included in the CALIFA final data \citep[DR3, ][]{2016A&A...594A..36S}.\\  

The details of the survey, sample, observational strategy, and data reduction are explained in \cite{2012A&A...538A...8S}. All galaxies were observed using the Postdam Multi Aperture Spectrograph \citep[PMAS; ][]{2005PASP..117..620R} in the PPAK configuration \citep{2004AN....325..151V,2006NewAR..50..355K,2006PASP..118..129K}, that is, a retrofitted bare fibre bundle IFU which expands the field-of-view (FoV) of PMAS to a hexagonal area with a footprint of $74\times65$ arcsec$^{2}$, which allows us to map the full optical extent of the galaxies up to two to three disk effective radii on average. This is possible because of the diameter selection of the sample \citep[hereafter W14]{2014A&A...569A...1W}. The observing strategy guarantees a complete coverage of the FoV, with a final spatial resolution of full width at half maximum (FWHM) $\sim$3", corresponding to $\sim$1 kpc at the average redshift of the survey. The sampled wavelength range and spectroscopic resolution (3745-7500 \AA, $\lambda$/$\bigtriangleup\lambda$ $\sim$850 for the low-resolution setup, that we use in this work) are more than sufficient to explore the most prominent ionized gas emission lines and to deblend and subtract the underlying stellar population \citep[e.g.,][]{2012A&A...538A...8S,2012A&A...540A..11K,2013A&A...557A..86C}. The dataset was reduced using version 1.5 of the CALIFA pipeline. The flux calibration, signal-to-noise ratio (S/N) and related uncertainties of the CALIFA data products have been thoroughly discussed in several articles of the CALIFA collaboration \citep[e.g., ][]{2012A&A...538A...8S,2014A&A...561A.130C,2015A&A...576A.135G}. For the 1st CALIFA Data Release \citep[DR1, ][]{2013A&A...549A..87H} the collaboration performed a data quality test showing that the sample reached a median limiting continuum sensitivity of 10$^{-18}$ erg s$^{-1}$ cm$^{-2}$ \AA$^{-1}$ arcsec$^{-2}$ at 5635 \AA, and 2.2 10$^{-18}$ erg s$^{-1}$ cm$^{-2}$ \AA$^{-1}$ arcsec$^{-2}$ at 4500 \AA, for the V500 and V1200 setup respectively, which corresponds to limiting r- and g- band surface brightnesses of 23.6 mag arcsec$^{-2}$ and 23.4 mag arcsec$^{-2}$, or an unresolved emission-line flux detection limit of roughly 10$^{-17}$ erg s$^{-1}$ cm$^{-2}$ arcsec$^{-2}$ and 0.6 10$^{-17}$ erg s$^{-1}$ cm$^{-2}$ arcsec$^{-2}$, respectively. The same limits, or slight improvements, were found in posterior data releases.

\section{Sample characterization}
\label{Sample characterization}

\subsection{Galaxy sample}
\label{Galaxy sample}

The starting point of this work were the 350 CALIFA galaxies observed using the low-resolution setup until September 2014. From this initial sample we selected the spiral galaxies and discarded ellipticals and lenticulars, that have no gas and do not host big processes of stellar formation. We also selected those spirals that are isolated, as our objective is analyzing \hh regions not affected by particular processes of interactions or mergers. Combining the isolated and merging classification and the morphological type designation from the CALIFA survey with the Hyperleda\footnote{http://leda.univ-lyon1.fr/} catalog \citep{2014A&A...570A..13M} classification as a matter of extra precaution, we finally selected 263 galaxies, that from now on constitute the main sample of our work.  

The main properties and characteristics of the 263 galaxies are included in a table that can be found in this paper online version. A part of this table is shown as an example in Table \ref{table_califa_galaxies}.

As it was important to ensure that the properties and parameter ranges of the 263 galaxies fulfilled the statistical properties of the whole CALIFA sample, with the only exception of the exclusion of the earlier morphological types, some of the main characteristics of the galaxies are particularly described in the following sections.

\subsubsection{Redshifts and distances}
\label{Redshifts and distances}

\begin{figure*}
\centering
\includegraphics[scale=0.6]{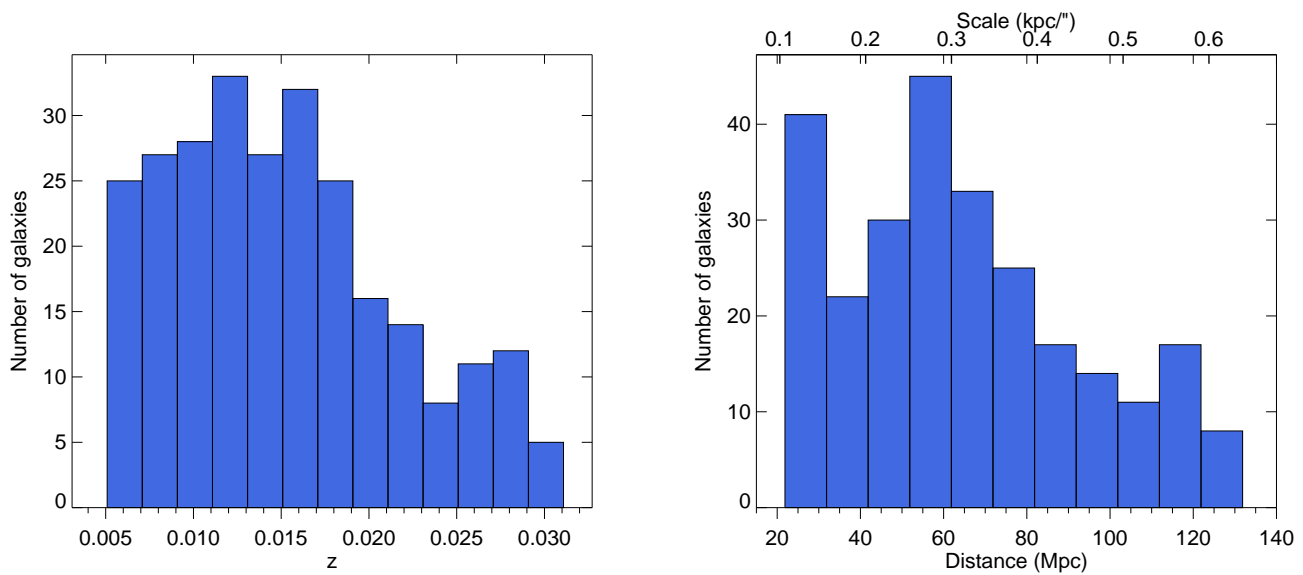}
\caption{Redshifts values (left) and of distance values in Mpc
  (right). Both histograms correspond to the galaxy sample of this work.}
\label{gal_z_distances}
\end{figure*}

Redshift values of our galaxies are those given by the CALIFA survey, that obtained them from the SIMBAD database on January 2010 (W14). Our galaxy sample has a redshift range, shown in Fig.\ref{gal_z_distances}, that covers the whole range of redshift values selected by CALIFA for its mother sample (0.005 < z < 0.03).

Distances for the CALIFA mother sample were obtained from NED and Hyperleda, finally adopting the NED-infall-corrected ones as their fiducial distances. In this work we adopt the distances calculated from the distance moduli given by Hyperleda, which are corrected  for Virgo-centric infall. The distance range for our galaxy sample is also included in Fig. \ref{gal_z_distances}, along with the scale range expresed in kpc/".

\subsubsection{Morphological classification}
\label{Morphological types}

We adopted the morphological classification performed by the CALIFA team (W14). The CALIFA collaboration found that the morphological classifications available from public databases were incomplete for the CALIFA sample \citep[e.g., Galaxy Zoo 2, 535 matches][]{2013MNRAS.435.2835W} or missing a consistent classification in Hubble subtypes (NED). Therefore they undertook their own reclassification, using human by-eye classification (see W14). 

\begin{figure}
\centering
\includegraphics[scale=0.5]{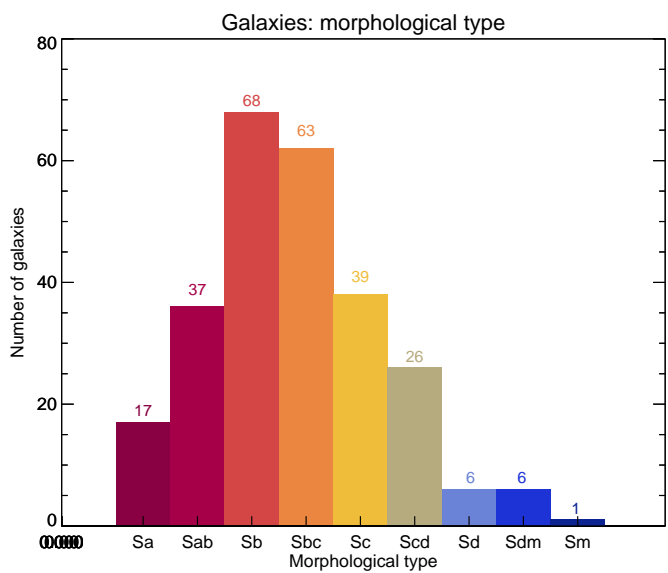}
\caption{Distribution of morphological types for the galaxy sample of this work.}
\label{gal_mtype}
\end{figure}

One of the defining characteristics of CALIFA mother sample is that it contains galaxies of all morphological types. In our case we only have spirals by selection, but our sample also comprises galaxies of all spiral morphological types. The morphological type histogram of our sample (Fig. \ref{gal_mtype}) follows a similar pattern in the Sa-Sm types that the one we can observe in the analogous histogram of the CALIFA mother sample (see W14). 

Regarding the presence of bars and rings, using the classification by Hyperleda we find a 40.3\% of barred galaxies in our sample, and a 17.1\% of galaxies where the presence of a ring can be observed. 

\subsubsection{Inclination}
\label{Inclination}

As described in W14, inclination may be the cause of a selection effect in the CALIFA mother sample, and thus in ours. Isophotal sizes of flattened, transparent (no attenuation) galaxies vary with inclination, due to the projected change of surface brightness \citep[e.g.,][]{1923Obs....46...51O}. It is therefore easier for an inclined disk galaxy to get into a sample defined by a minimum apparent isophotal size than it is for a face-on system of the same intrinsic dimensions. The magnitude of this effect depends on the degree of transparency; it is strongest for a fully transparent galaxy, and it disappears when the system is opaque, so that only its surface is observed. Therefore it can be expected to find an excess of galaxies with high inclinations in the CALIFA sample, at least among disk-dominated systems, and this may happen also in the spiral sample of this work. 

During the characterization of their mother sample, the CALIFA team detected this effect when studying isophotal major and minor axes delivered by the SDSS photometric pipeline, that can be combined into an axis ratio at the outer 25 mag/arcsec$^{2}$ level (see W14). They found that the histogram of isophotal axis was clearly skewed toward low values of b/a, providing an indication of the considered selection effect. Furthermore, they also represented the 55 galaxies of the CALIFA mother sample that have Mr > -18.6, that is, that are below the completeness limit. Nearly all of these galaxies have axis ratios below 0.4, and it can be visually confirmed that these are predominantly disk-dominated systems that are close to edge-on. The CALIFA team presumed that very few, if any, of these galaxies would have been included into the CALIFA sample of seen face-on, their angular sizes have been boosted through inclination, just enough to promote them into the sample. They reached the conclusion that while the CALIFA sample has a higher proportion of inclined disk galaxies at the faint end, the overall effect is not large. Specifically for the galaxies close to and below the low-luminosity completeness limit there is at any rate a clear surplus of galaxies with very high inclinations in the CALIFA sample.  

\begin{figure}
\centering
\includegraphics[scale=0.5]{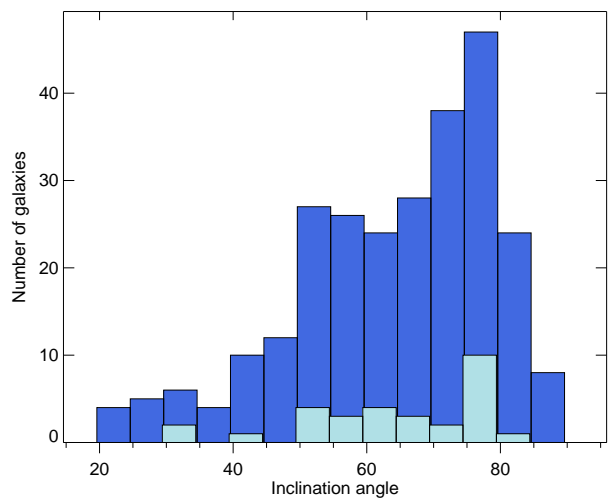}
\caption{Inclination values for this work galaxy sample, calculated from the observed axis ratios. Values of all galaxies are represented in dark blue, and overplotted in light blue are the values of the galaxies with \textit{M}r > -18.6.}
\label{gal_incl}
\end{figure}

We derive the inclination values for our galaxy sample from the b/a axis ratios given by CALIFA, that obtained them by calculating light moments. The final b/a value is the mean of the axis ratios of ellipses containing 50\% and 90\% of the total flux (see W14). To obtain the final inclination values we use the expression given by \cite{1958MeLuS.136....1H}, where the value of the axial ratio for an edge-on system parameter as a function of the galaxy morphological type is given by \cite{1972MmRAS..75...85H}. In Fig. \ref{gal_incl} we represent the inclination values for the whole sample and also for the 31 galaxies that are below the CALIFA completeness limit. We find a distribution skewed toward high inclination values, that is more prominent for the faint galaxies. We consider therefore that we are detecting the selection effect already existing in the CALIFA mother sample, and that specifically affects galaxies with low luminosity.

This high number of high inclination or edge-on galaxies have to be taken into account, as it implies that these galaxies will have higher uncertainties in the determination of the distances of their star forming regions to the center of the system, as well as it can affect the morphological classification and other factors.

\subsubsection{Effective radius}
\label{Effective radius}

In this work we used the disk effective radius, classically defined as the radius at which one half of the total light of the system is emitted, as the factor of normalization to analyse the galaxy properties' radial distributions and compared them galaxy to galaxy. Concerning the study of radial gradients and 2D distribution of galactic properties, although there are a high number of studies about the issue, we find a large degree of discrepancy among them. One of the factors that may cause these differences is the fact that it does not exist an uniform method to analyse the gradients. In some cases the physical scales of the galaxies (i.e., the radii in kpc) are used \citep[e.g.,][]{2012ApJ...754...61M}. In others the scale-lengths are normalized to the R$_{25}$ radius, that is, the radius at which the surface brightness in the B band reach the value of 25 mag/arcsec$^{2}$ \citep[e.g.,][]{2011MNRAS.415.2439R}. Finally, a reduced number of studies try to normalize the scale-length based on the effective radii. \cite{1989epg..conf..377D} already showed that the effective radius seems to be the best to normalize the abundance gradients. Using the physical scale of the radial distance or the normalized one to an absolute parameter like the R$_{25}$ radius does not produce gradients that we can compare galaxy to galaxy, since in both cases the derived gradient is correlated with either the scale-length of the galaxy or its absolute luminosity.

\begin{figure}
\centering
\includegraphics[scale=0.5]{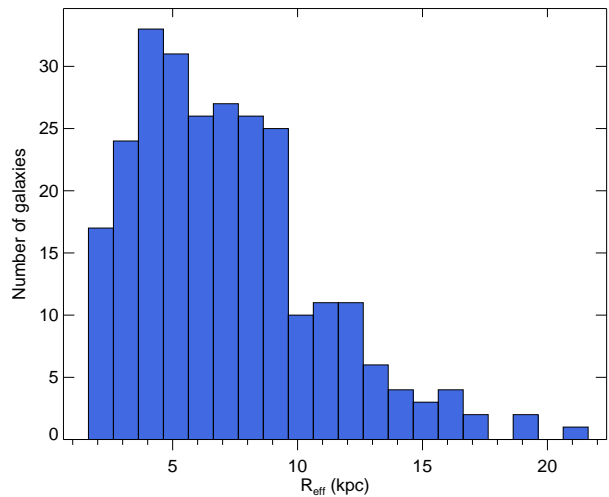}
\caption{Effective radius values (kpc) of this work galaxy sample.}
\label{gal_reff}
\end{figure}

We used the effective radii values estimated by the CALIFA survey, whose
calculation is described in \cite{2012A&A...546A...2S}. It is based on an
analysis of the azimuthal surface brightness profile, derived from an elliptical
isophotal fitting of the ancillary g-band images collected for the galaxies
\citep[extracted from the SDSS imaging
  survey,][]{2000AJ....120.1579Y,2011A&A...534A...8M}. When these ancillary
images were not available, the B band was used \citep{2011A&A...534A...8M}.
Our galaxy sample contains a wide range of effective radii, as shown in
Fig. \ref{gal_reff}, wich implies that we selected galaxies with a wide
range of sizes.

\subsubsection{Color-magnitude diagrams}
\label{Color-magnitude diagrams}

\begin{figure}
\centering
\includegraphics[scale=0.45]{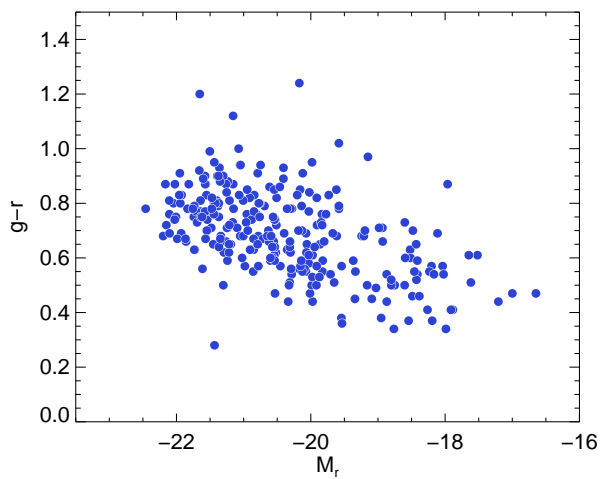}
\caption{Distribution of this work galaxy sample in the g-r vs. $\textit{M}_{r}$ color-magnitude diagram. }
\label{gal_cmdiagram}
\end{figure}

The color-magnitude diagram of the 263 galaxies of our sample, represented in Fig. \ref{gal_cmdiagram}, show that they fully cover the range in absolute magnitudes where the CALIFA sample is representative of the overall galaxy population. This is consistent with the results obtained by \cite{2014MNRAS.440..889S}, that signaled the fact that late-type galaxies do not separate into a blue cloud and a red sequence, but rather span almost the entire color range without any gap or valley.

\subsection{Spectroscopic information of the \hii\ regions sample}
\label{HII regions sample}

The \hh region segregation and their corresponding spectra extraction is
performed using a semi-automatic procedure named {\sc HIIexplorer}, described in 
\citet{2012A&A...546A...2S} and \citet{2012ApJ...756L..31R}. It is based on the
assumptions that: (a) \hh regions are peaky and isolated structures with a strong
ionized gas emission, which is significantly above the stellar continuum
emission and the average ionized gas emission across the galaxy. This is
particularly true for H$\alpha$ because (b) \hh regions have a typical physical
size of about a hundred or a few hundred parsecs
\citep[e.g.,][]{1997ApJS..108..199G,2011ApJ...731...91L,2003AJ....126.2317O},
which corresponds to a typical projected size of a few arcsec at the distance of
the galaxies. These basic assumptions are based on the fact that most of the
H$\alpha$ luminosity observed in spiral and irregular galaxies is a direct
tracer of the ionization of the interstellar medium (ISM) by the ultraviolet
(UV) radiation produced by young high-mass OB stars. Since only high-mass,
short-lived stars contribute significantly to the integrated ionizing flux, this
luminosity is a direct tracer of the current star formation rate (SFR),
independent of the previous star formation history. Therefore, clumpy structures
detected in the H$\alpha$ intensity maps are most probably associated with
classical \hh regions (i.e., those regions for which the oxygen abundances have
been calibrated).

For each region selected by {\sc HIIexplorer}, we extracted a integrated spectrum of the spaxels belonging to that region. For each individual extracted spectrum we then modeled the stellar continuum using {\sc FIT3D}, a fitting package described in \cite{2006AN....327..167S} and \cite{2011MNRAS.410..313S}. The {\sc FIT3D} version used at the moment of this fitting adopted a simple SSP template grid with 12 individual populations. It comprises four stellar ages (0.09, 0.45, 1.00, and 17.78 Gyr), two young and two old ones, and three metallicities (0.0004, 0.019, and 0.03, that is, subsolar, solar, or supersolar, respectively). The models were extracted from the SSP template library provided by the MILES project \citep{2010MNRAS.404.1639V,2011A&A...532A..95F}. The \cite{1989ApJ...345..245C} law for the stellar dust attenuation with an specific attenuation of R$_{V}$ = 3.1 was adopted, assuming a simple screen distribution. 

Individual emission line fluxes were measured in the stellar-population subtracted spectra performing a multicomponent fitting using a single Gaussian function. The equivalent widths for each \hh region and line were estimated using the results from the fitting analysis instead of the classical procedure, by dividing the emission line integrated intensities by the underlying continuum flux density. The continuum was estimated as the median intensity in a bandwidth of 100 \AA, centered in the line, using the gas-subtracted spectra provided by the fitting procedure. For further details about both processes, see \cite{2012A&A...546A...2S}. The errors in the determination of the emission line fluxes and their reliability are discussed extensively in \cite{2016RMxAA..52...21S} and \cite{2016RMxAA..52..171S} for the {\sc Pipe3D/FIT3D} fitting technique.

After applying all the process to the 263 CALIFA galaxies datacubes, we detected
a total of 12891 \hh regions. Nevertheless, not all these regions can be
accepted as confirmed \hh regions due to the high level of noise of some spectra
or the non-physical values of some parameters such as H$\alpha$/H$\beta$
. Therefore we apply a quality control process, considering the
following criteria to ensure that we are working with physical bona fide \hh regions and
avoid selection uncertainties: (i) EW(H$\alpha$) > 6\AA, following
\cite{2010MNRAS.403.1036C, 2015A&A...574A..47S}. (ii) H$\alpha$/H$\beta$ >
2.7. We consider the theoretical value for the intrinsic line ratio
H$\alpha$/H$\beta$ from \cite{2006agna.book.....O}, assuming case B
recombination (optically thick in all the Lyman lines), an electron density of
$n_{e}$=100 cm$^{-3}$ and an electron temperature of $T_{e}$=10$^{4}$ K. Lowering the electron temperature to 5000K, keeping the electron density constant, increases the Balmer decrement H$\alpha$/H$\beta$ by a factor of 1.05 and translates to an uncertainty of 0.04dex in c(H$\beta$) for the reddening curve employed. We also have
included a certain margin to account for uncertainties in the observational
values of the emission lines. (iii) H$\alpha$/H$\beta$ < 6. This value corresponds to an extinction of
$\sim$2.3 mag. We consider that beyond this point values are not physical. (iv)
F(H$\beta$) > 0.5 10$^{-16}$ erg s$^{-1}$ cm$^{-2}$. To avoid lines with very
small S/N (v) F(\oiii~$\lambda$5007) > 0.5 10$^{-16}$ erg s$^{-1}$
cm$^{-2}$. This condition was introduced due to non-physical values of the
\oiii~$\lambda$5007/H$\beta$ emission-line ratio observed for some regions in a
preliminar version of the data. We finally obtained a sample of 9281 selected \hh
regions.

\begin{figure*}
\centering
\includegraphics[scale=0.5]{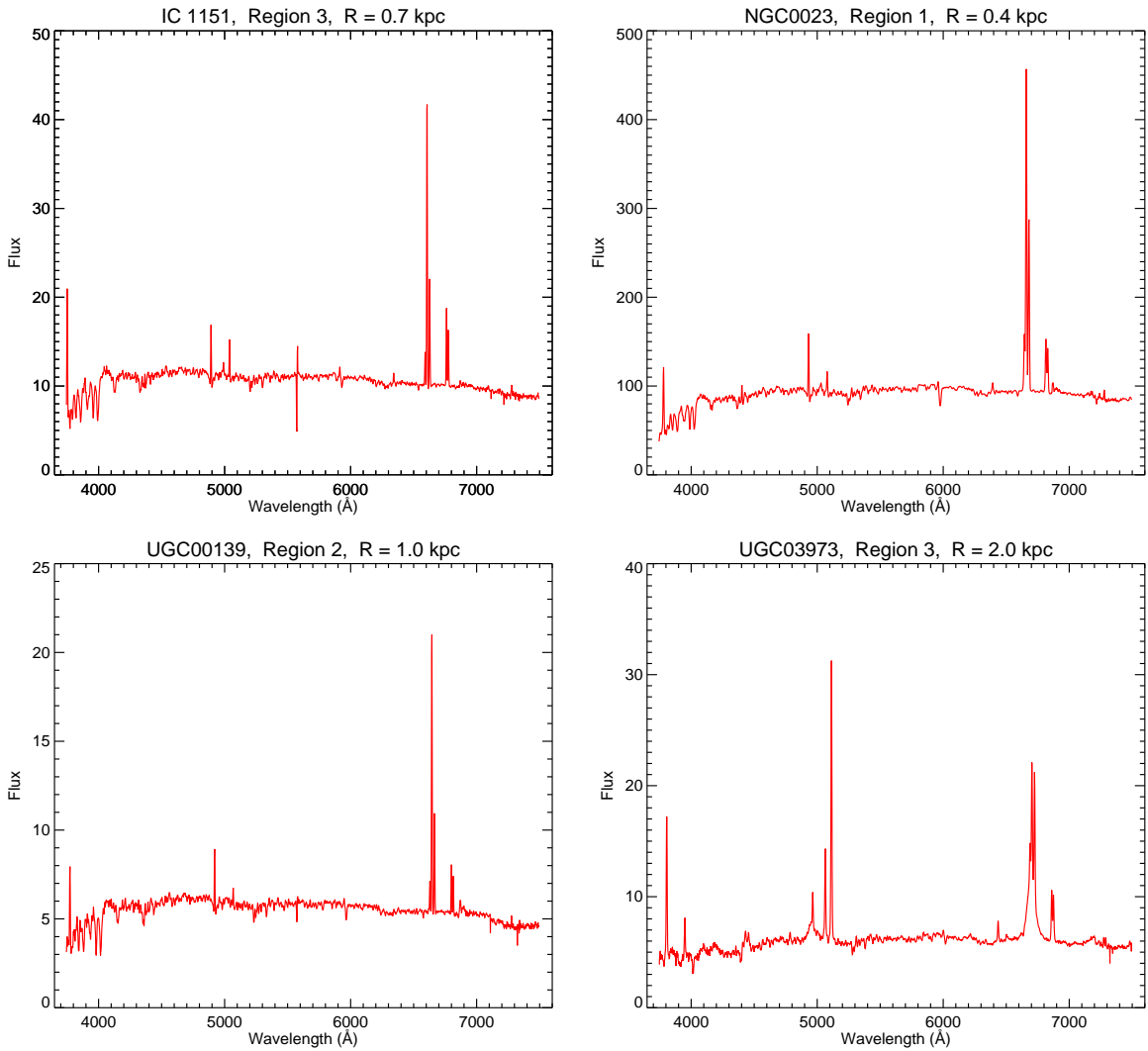}
\caption{Some of the inner regions extracted spectra, before the subtraction of the stellar continuum. Galaxy name, ID of the region in the galaxy and distance to the center are shown in the titles. Flux is expressed in units of 10$^{-16}$ erg s$^{-1}$ cm$^{-2}$.}
\label{inner_spectra}
\end{figure*}

\begin{figure*}
\centering
\includegraphics[scale=0.5]{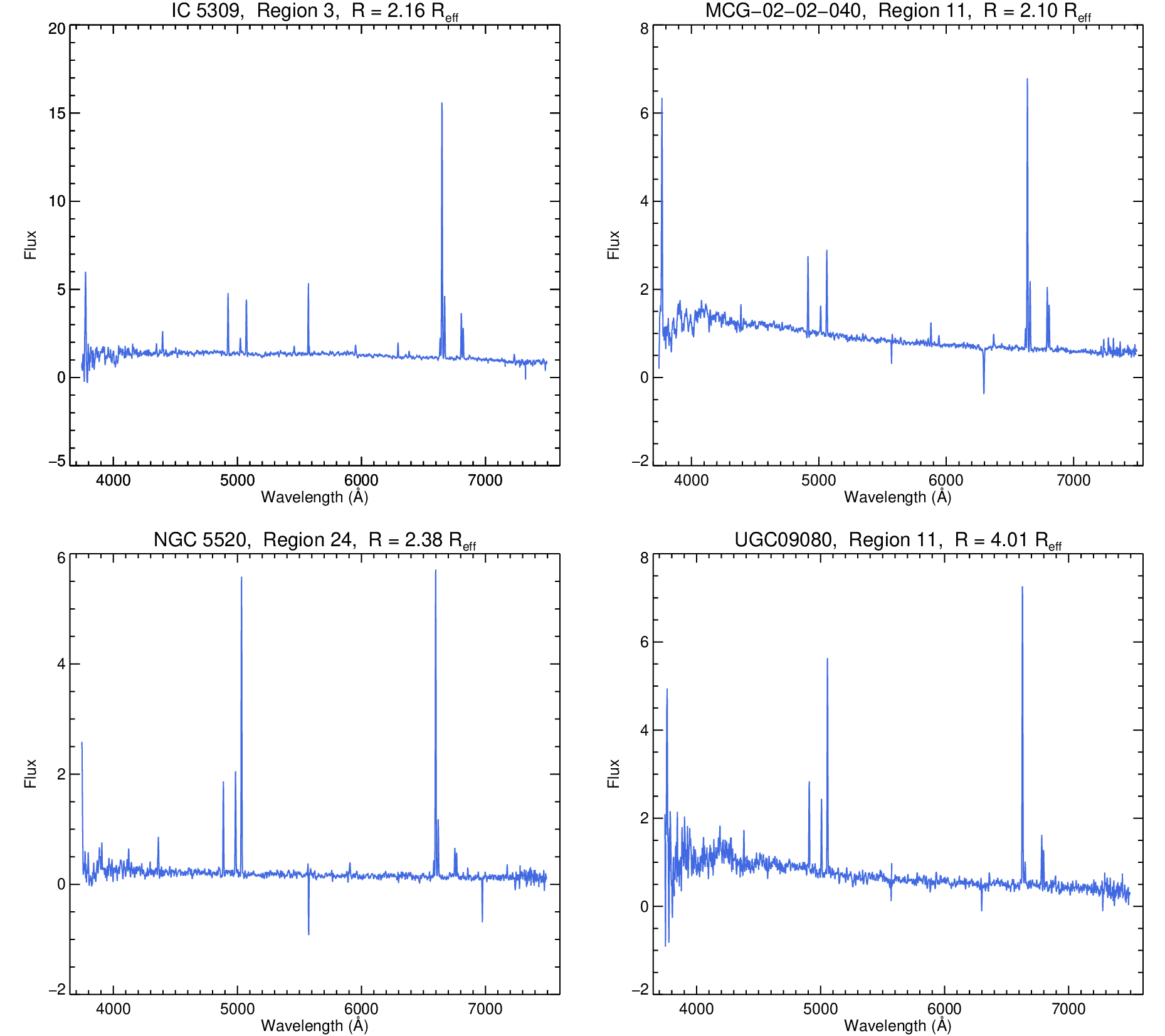}
\caption{Some of the outer regions extracted spectra, before the subtraction of the stellar continuum. Galaxy name, ID of the region in the galaxy and distance to the center are shown in the titles. Flux is expressed in units of 10$^{-16}$ erg s$^{-1}$ cm$^{-2}$.}
\label{outer_spectra}
\end{figure*}

\subsubsection{Inner regions sample}
\label{Inner regions sample}

We considered as inner regions those that fulfil the criterium established by \cite{2015MNRAS.451.3173A}, based on the observed separation between nuclear and disk region rings as a function of the galaxy luminosity. Following this criterium, inner regions are those located closer to the center than the distance defined by the expression:
\begin{equation}
\log {\rm R(kpc)} = -0.204\cdot{\rm M_B} - 3.5 
.\end{equation}

We calculated the B-band magnitude from the g and r magnitudes from SDSS, using
the transformation given by Lupton (2005)\footnote{http://www.sdss.org/dr12/algorithms/sdssubvritransform/}. After applying this
criterium to the primary regions sample we obtained a total of 794 inner
regions. Nevertheless, by detailed examination of the region extracted spectra
we note that not all the regions have clear visible emission from other typical
\hh region spectral lines, apart from H$\alpha$. This could be previously
expected, due to the strong stellar continuum found in the inner part of the
galaxies, which prevents the detection of weaker spectral lines. Therefore we made a second selection process by examination by-eye, discarding those regions whose spectra were dominated by stellar continua with the presence of only weak Halpha emission, or by some gas emission features not clearly detectable. After that we obtained a final sample of 725 regions with spectra where H$\alpha$,
H$\beta$, \oiii~$\lambda$5007, \nii~$\lambda\lambda$6548,6583 or the
\sii~$\lambda$6717,6731 doublet are measurable.

\subsubsection{Outer regions sample}
\label{Outer regions sample}

We considered as external \hh regions those that are located at a distance larger
than two effective radii ($\textit{R}_{eff}$) from the center of the galaxy. It is
around this radius where a certain amount of flatness is found in abundance
gradients in spirals \citep{1989epg..conf..377D,2014A&A...563A..49S,2016A&A...585A..47M,2016ApJ...830L..40S}. From the
primary sample of 9281 regions obtained, a total of 1027 regions were located
beyond the considered distance to the center of the system. Nevertheless, as happened with the inner regions, not all of them show \hh region emission
features or do not have a high enough S/N. We therefore applied a second selection by eye,
obtaining a final sample of 671 inner regions.

\section{Results and discussion}
\label{Results and discussion}

\subsection{Observational and functional parameters}
\label{Observational and functional parameters}

After the process of extraction and selection we get a final sample of 725 inner regions and 671 outer regions. Some of the inner regions spectra are shown as an example in Fig. \ref{inner_spectra}, while some of the outer region spectra are shown in Fig. \ref{outer_spectra}.

The number of inner and outer regions included in the final samples allows us to develop a statistical analysis of some spectroscopic properties, especially those based on the the strongest detected emission lines, such as the following: 

(i) EW(H$\alpha$), the equivalent width of H$\alpha$, which is directly related to the fraction of very young stars in the region 

(ii) $A_V$, the dust attenuation, calculated using the Balmer decrement according to the reddening function of \cite{1989ApJ...345..245C}, assuming $R \equiv A_V /E(B - V ) = 3.1$. Theoretical value for the intrinsic line ratio H$\alpha$/H$\beta$ was considered as explained in Sect. \ref{HII regions sample} 

(iii) L(H$\alpha$), the H$\alpha$ luminosity, obtained from the reddening-corrected H$\alpha$ flux, considering the distances to the corresponding galaxies 

(iv) \nii~$\lambda$6583/H$\alpha$ line ratio, related to the oxygen abundance of the ionized gas, and that along with the \oiii~$\lambda$5007/H$\beta$ provides information about the nature of the ionization source of the region 

(v) \oii~$\lambda$3727/\oiii~$\lambda$5007 line ratio, related to the ionization parameter log \textit{u}, a measurement of the strength of the ionization radiation \citep{2000MNRAS.318..462D} 

(vi) \sii~$\lambda$6717/\sii~$\lambda$6731 line ratio, related to the electron density ($n_{e}$) of the ionized gas.

\begin{figure*}[h!]
\centering
\includegraphics[scale=0.52]{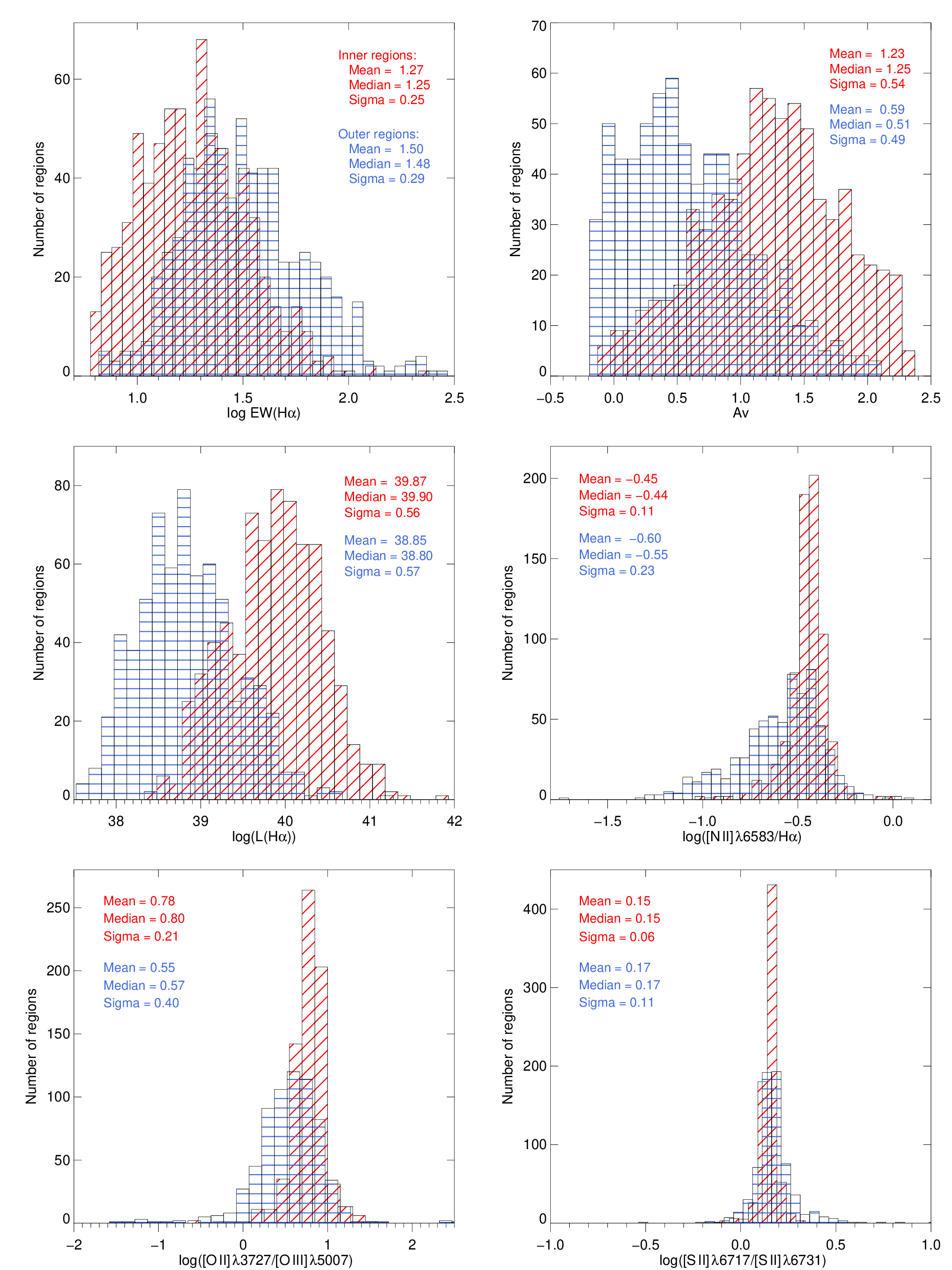}
\caption{From top to bottom and left to right: Equivalent width of H$\alpha$, dust attenuation, H$\alpha$ luminosity and of the \nii~$\lambda$6583/H$\alpha$, \oii~$\lambda$3727/\oiii~$\lambda$5007, and \sii~$\lambda$6717/\sii~$\lambda$6731 emission-line ratios for the inner (red diagonally-hatched diagram) and outer (blue horizontally-hatched diagram) region samples.}
\label{regions_hist_joint}
\end{figure*}

Figure \ref{regions_hist_joint} shows EW(H$\alpha$), $A_{V}$, L(H$\alpha$) and \nii~$\lambda$6583/H$\alpha$, \oii~$\lambda$3727/\oiii~$\lambda$5007, and \sii~$\lambda$6717/\sii~$\lambda$6731 emission-line ratio histograms for inner and outer region samples. Different trends and average values can be observed for both samples. 

Firstly, EW(H$\alpha$) histograms show smaller EW(H$\alpha$) values for inner \hh regions. \cite{2014A&A...563A..49S}, that work with a sample of 7016 \hh regions from 227 CALIFA galaxies also selected and extracted with {\sc HIIexplorer}, finds a strong log-linear correlation between EW(H$\alpha$) and the percentage of young stars in the regions, obtained from the {\sc FIT3D} fitting of the underlying stellar population. This correlation is valid for regions with EW(H$\alpha$) > $6\AA$ and with a percentage of young stars over 20\%. All our regions have EW(H$\alpha$) over 6\AA, as it is one of our selection criteria. We consider the smaller values of EW(H$\alpha$) for our inner regions sample to be caused by the greater influence of the underlying stellar populations in those regions, and therefore to smaller percentages of young population (see Sect. \ref{Ionizing and photometric masses}). Secondly, on the middle histograms we observe larger $A_{V}$ values for inner regions, denoting greater dust attenuation. Finally, L(H$\alpha$) histograms reveal larger luminosities for inner regions. This is concordant with previous observations of very luminous \hh regions located close to their galactic nucleus \citep{2015MNRAS.451.3173A}, although it may be also influenced by a selection bias causing that in central regions, where the underlying continuum have a great influence, only the more luminous \hh regions can be detected. This possible selection bias will be studied afterwards in this section.

\begin{figure*}
\centering
\includegraphics[scale=0.5]{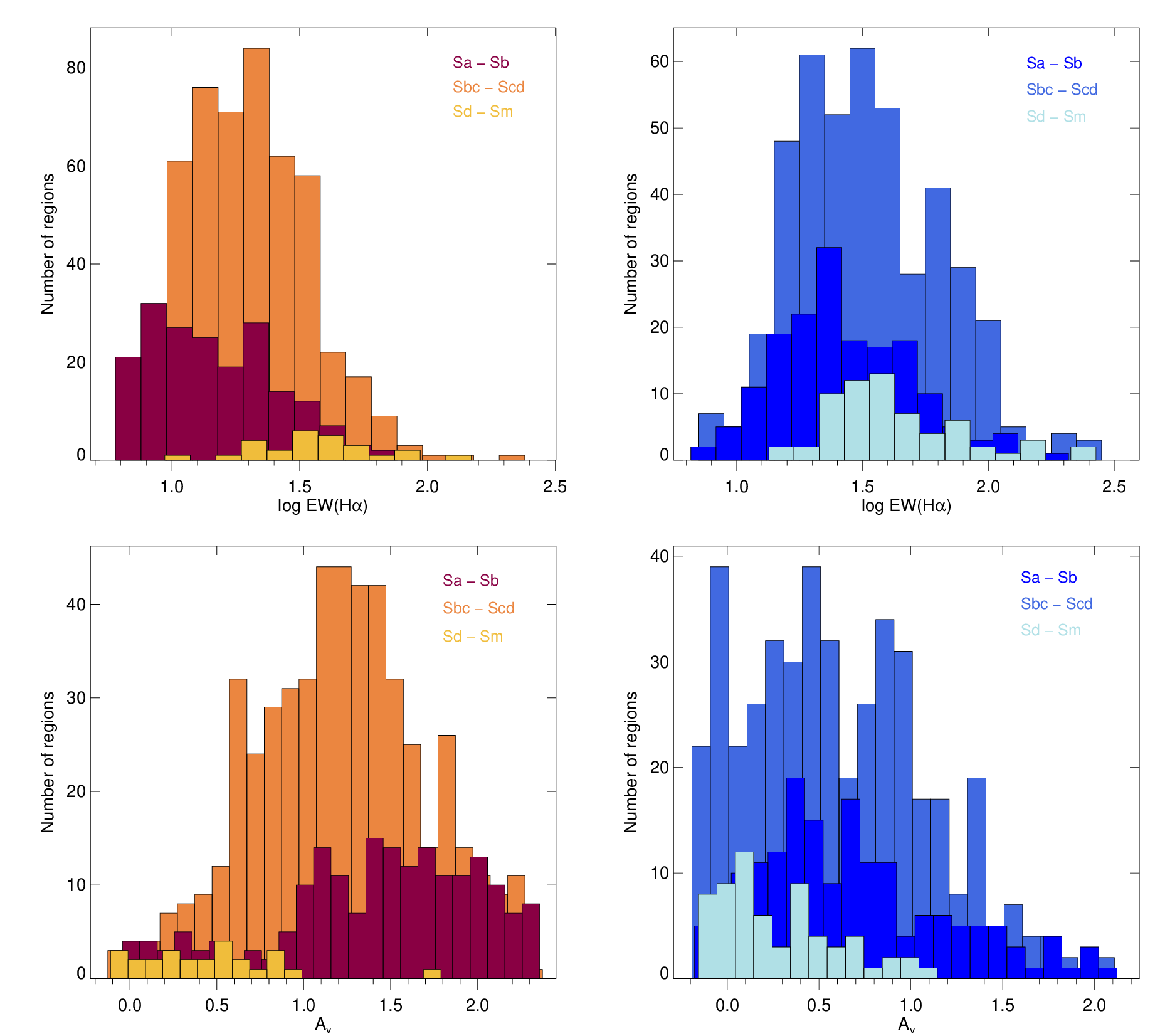}
\caption{Left column: EW(H$\alpha$) and $A_{V}$ histograms from the inner \hh regions sample as a function of their galaxies morphological types. Regions with Sa, Sab and Sb type galaxies are colored in dark red, those with Sbc, Sc and Scd type galaxies in orange and those with Sd, Sdm and Sm type galaxies are colored in gold. EW(H$\alpha$) is expressed in units of \AA and $A_{V}$ in magnitudes. Right column: Histograms with the same parameters as those from the outer \hh regions sample. Regions with Sa, Sab and Sb type galaxies are colored in dark blue, those with Sbc, Sc and Scd type galaxies in blue and those with Sd, Sdm and Sm type galaxies are colored in light blue.}
\label{regions_hist_EW_Av_mtype}
\end{figure*}

We find differences between the equivalent width and extinction values of inner regions as a function of the morphological type of the galaxies they belong to, as can be seen in Fig. \ref{regions_hist_EW_Av_mtype}. In early type spiral galaxies the greater prominence of the galaxy bulge implies greater influence of older underlying population, which means a decrease of the ionizing population percentage and of the equivalent width values. It also implies higher amounts of dust and therefore higher extinction values for these regions. On the contrary, late-type spirals, with little to none bulge component, have increasingly higher equivalent width values and less extinction. For the outer regions, differences between different morphological types are almost negligible.

Clear differences between inner and outer regions are also detected in histograms representing emission-line ratios, in
Fig. \ref{regions_hist_joint}. In the \nii~$\lambda$6583/H$\alpha$ histograms, at the top of the figure, we observe greater values for the inner regions. This emission-line ratio is related with the oxygen abundance, which is therefore
higher in the inner regions, as could be expected for more evolved stellar populations and more enriched interstellar medium. Inner regions also have greater values of \oii~$\lambda$3727/\oiii~$\lambda$5007 ratio, indicating in this case smaller values of the ionization parameter. In the case of \sii~$\lambda$6717/\sii~$\lambda$6731 histograms we find similar average values for outer and inner regions, corresponding to electron density values smaller than 10 cm$^{-3}$ \citep{2006agna.book.....O}. All the emission-line ratio histograms show sharper distributions for inner regions, while the outer regions are more scattered. This may be caused by lower uncertainties in the inner regions, whose spectra have higher S/N values.

\begin{figure}
\centering
\includegraphics[scale=0.48]{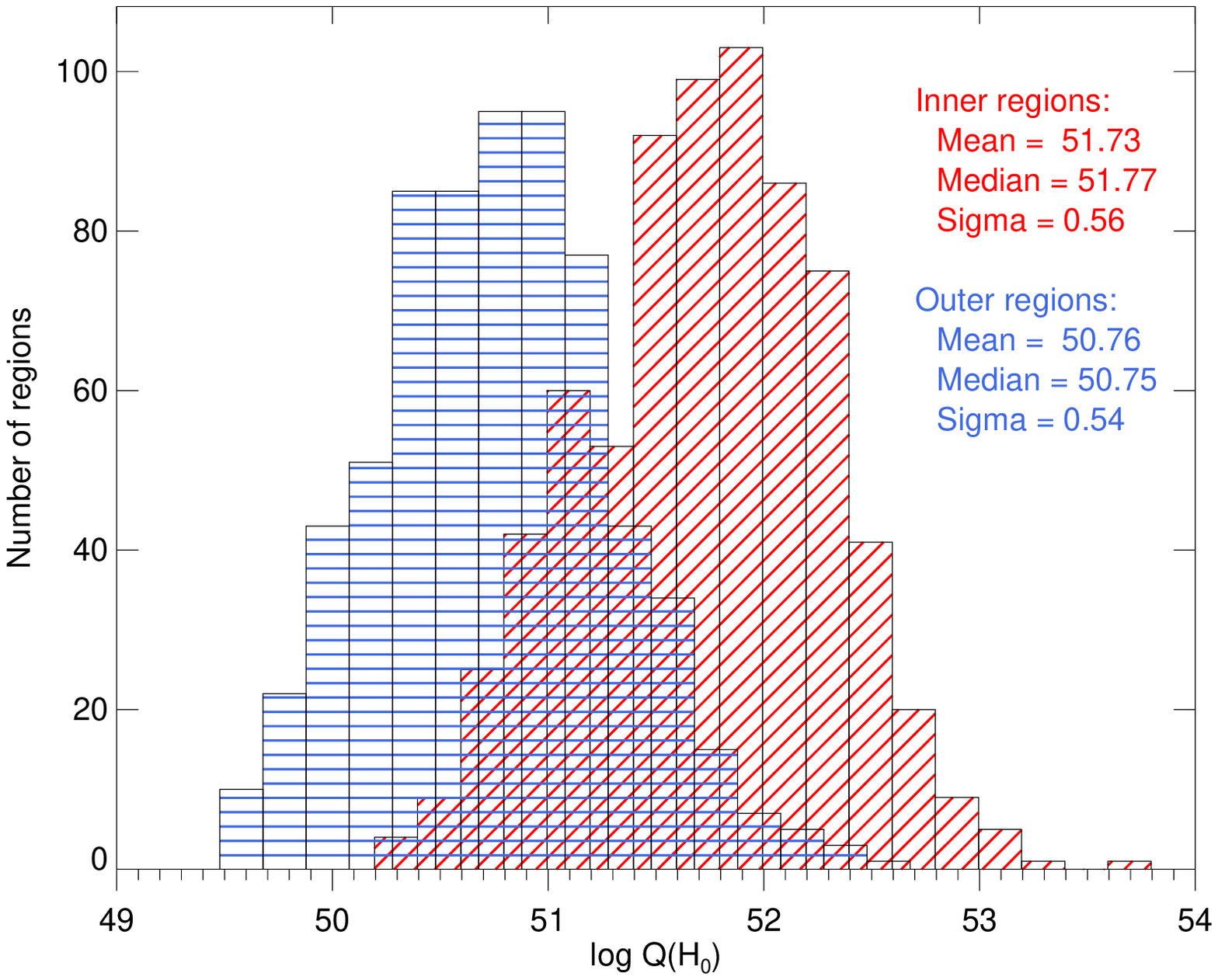}
\caption{Number of ionizing photons for the inner (red diagonally-hatched diagram) and outer (blue horizontally-hatched diagram) region samples.}
\label{hist_ionizingphotons}
\end{figure}

\begin{figure}
\centering
\includegraphics[scale=0.48]{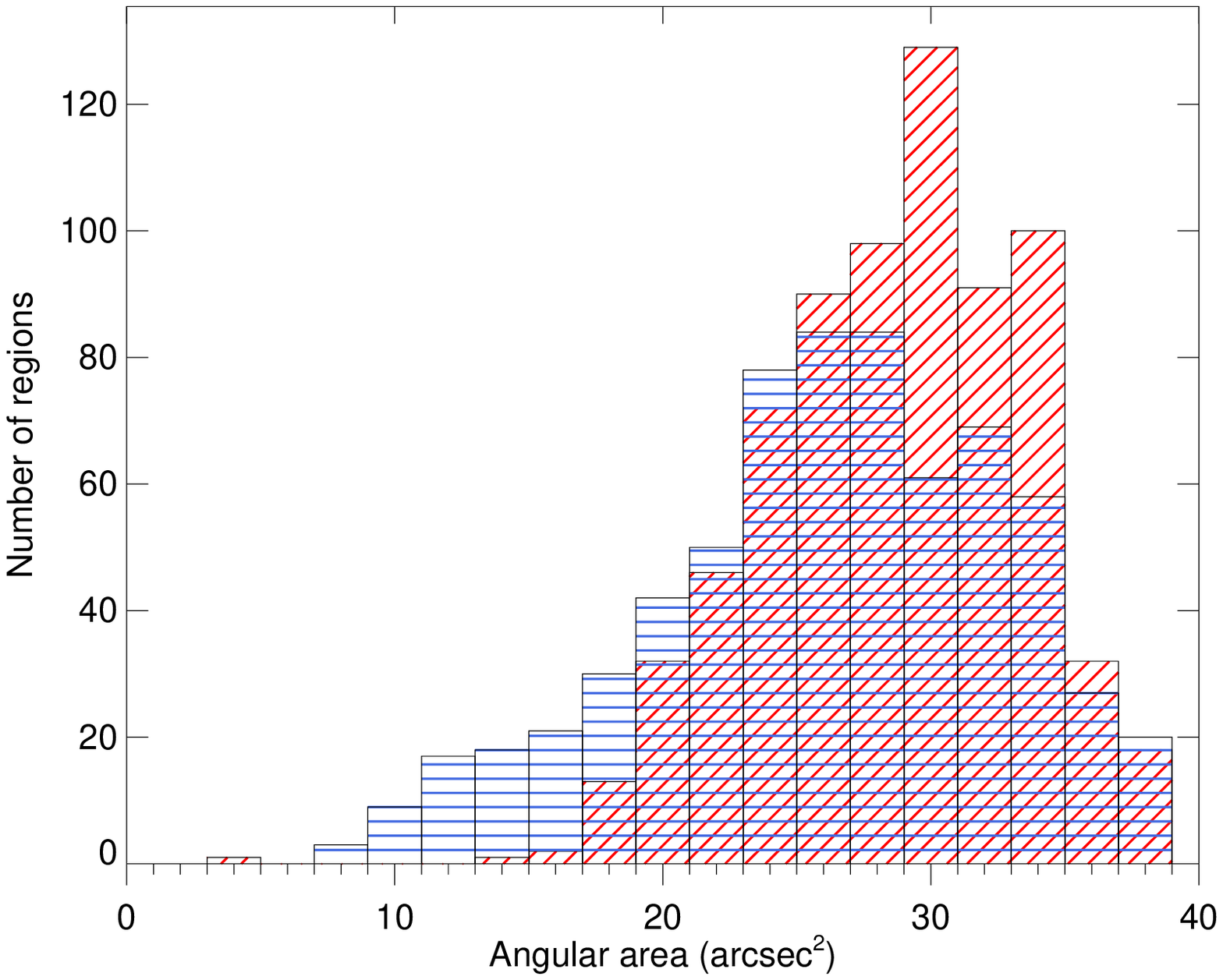}
\caption{Angular area in arcsec$^{2}$ for the inner (red diagonally-hatched diagram) and outer (blue horizontally-hatched diagram) region samples.}
\label{hist_angulararea}
\end{figure}

Using the relation between the H$\alpha$ luminosity and the number of ionizing Lyman continuum photons given by \cite{1995ApJ...439..604G} we calculate the number of ionizing photons for every \hh region. Results obtained for both inner and outer samples are shown in Fig. \ref{hist_ionizingphotons}. Inner regions have higher values, as could be expected from their higher H$\alpha$ luminosity values. As we have mentioned, this could be an intrinsic property or the result of a selection bias, caused by the fact the smaller inner regions are not detected due to the lack of spatial resolution and/or contrast with respect to the bright bulge. In order to study the magnitude of this possible bias we represent the histograms of the angular area in arcsec$^{2}$ of the inner and outer regions, which are included in Fig. \ref{hist_angulararea}. We can see that although outer regions do, in fact, have a tail of smaller regions that is not present in the inner regions histogram (which is probably caused by this bias), the number of regions included in this tail is not enough to cause the difference of values ranges observed in the H$\alpha$ luminosity and the number of ionizing photons histograms. Therefore we conclude that, although this selection bias has a small influence, there is an intrinsic difference of luminosity and number of ionizing photons between inner and outer \hh regions.

\subsection{Systematics in diagnostic diagrams}
\label{Disc:Diagnostic diagrams}

Emission-line diagnostic diagrams \citep[introduced by][hereafter BPT]{1981PASP...93....5B} are a powerful way to study the nature of the dominant ionizing sources and changes in the physical conditions of the ionized nebulae, either from galaxy to galaxy, within each galaxy or within a particular nebulae. BPT diagrams work by exploring the location of certain line ratios, involving several strong emission lines with a dependence on the ionization degree and, to a lesser extent, on temperature or abundance. Through the application of different classification criteria \citep{2001ApJ...556..121K,2003MNRAS.346.1055K} diagnostic diagrams allow the separation of galaxies or galaxy regions into those dominated by ongoing star formation and the ones dominated by non-stellar processes.

The distributions of CALIFA inner region sample and outer region sample in one of the most classical BPT diagrams, the \oiii~$\lambda$5007/H$\beta$ vs \nii~$\lambda$6583/H$\alpha$ diagram, are displayed in Fig. \ref{BPT_figure}. In the case of the outer regions sample, data is color-coded according to their distance to the center of the galaxy bins that are indicated in the plot. It can be observed that the inner regions are mostly located to the bottom-right corner of the classical star forming branch, with the exception of a few regions located in the active galactic nuclei (AGN) zone demarcated by the \cite{2001ApJ...556..121K} classification line, that will be analyzed afterwards. On the contrary, outer regions are more distributed along the star forming branch, and are generally located closer to the top-left corner. This is not surprising, as regions closer to the center of the galaxies are expected to have higher metallicities. Furthermore, \cite{2014A&A...563A..49S} and \cite{2015A&A...574A..47S} found a clear correlation reflected in their \hh regions distribution on the BPT diagrams, relating lower percentages of young stars with higher values of \nii~$\lambda$6583/H$\alpha$. This is consistent with our results, as we would expect a larger contribution of the underlying stellar populations in the inner regions, and then lower percentage of young stars than in the outer regions. 

We have studied the spectra of the \hh regions that, according to \cite{2001ApJ...556..121K} classification criteria, are located in the AGN region of the BPT diagram in Fig. \ref{BPT_figure}. In the case of the outer regions sample, we have five regions located in AGN zone, belonging to the galaxies IC2101 (2 regions), IC1528, UGC09542 and UGC 09598. All the spectra have low S/N, with specifically very low values of the \oiii~$\lambda$5007 emission line. Considering this and the magnitude of the errors associated, we consider that the location of those regions in the AGN zone is due to the uncertainties associated to the emission line flux measurements. 

In the case of the inner regions sample, we have six regions located in the AGN zone. Three of them are close to the classification line, while the other three are far inside the AGN region. These last three regions belong to the galaxies NGC2410 and UGC03973 (two regions). NGC2410 is classified as a Seyfert 2 by NED \citep{2006A&A...455..773V} and UGC 03973 is classified as a Seyfert 1 by NED \citep{1998A&AS..130..285C}. Therefore we consider that the location of these regions in the AGN zone is due to their corresponding nucleus emission influence, and in fact the active nucleus emission features are easily detectable in their spectra. On the other hand, the other three regions belong to the galaxies IC2247, UGC00005 and UGC03151, that are not classified as active. Their spectra have low S/N values and, as explained for the outer regions case, we consider that their location in the AGN regions is due to the uncertainties in the emission line measurements.

\subsubsection{Observations of \hh regions with higher-spatial resolution}
\label{PINGS_sec}

During the analysis of our CALIFA \hh regions sample we considered the
possibility of including other IFS observations of \hh regions with
higher-spatial resolution such as the sample extracted from the PPAK IFS Nearby
Galaxy Survey \citep[PINGS;][]{2010MNRAS.405..735R}, in order to extend our data. 
While CALIFA galaxies have a redshift range of 0.005 < z < 0.03 (see
Sect. \ref{Redshifts and distances}),
PINGS galaxies have much lower redshifts. This implies a loss of resolution that
was studied by \cite{2014A&A...561A.129M} using some of the PINGS galaxies, that
were simulated at higher redshifts to match the characteristics and resolution
the galaxies observed by the CALIFA survey. Regarding the \hh region selection,
the authors conclude that at z $\sim$0.02 the \hh clumps can contain on average
between one and six of the \hh regions obtained from the original data at z $\sim$
0.001. This prevents a complete combined analysis of CALIFA and PINGS regions, as
parameters depending on the \hh regions size (luminosity, masses) are not
comparable. Despite that, we can consider PINGS regions when analyzing
properties where emission-line ratios are involved, that are independent of the
size of the regions.

From a total sample of 17 nearby spiral galaxies included in the PINGS galaxy
sample, we consider for this work those that are not involved in interaction or
merging processes, as we did for the CALIFA galaxies. Our PINGS galaxy sample is
therefore composed by four galaxies: NGC\,628, NGC\,1058, NGC\,1637 and
NGC\,3184. Table \ref{table_pings_galaxies} includes the main properties and
characteristics of these four galaxies.

\begin{figure*}
\centering
\includegraphics[scale=0.55, bb=36 180 886 548]{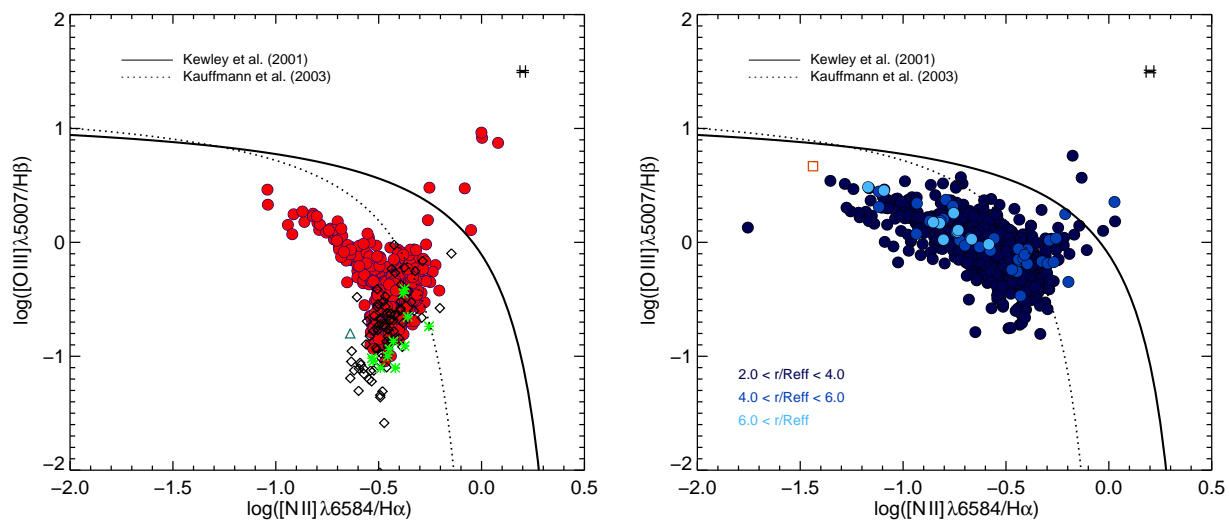}
\caption{Left: Diagnostic diagram of the CALIFA inner region sample (red circles). Data from PINGS inner region sample (black diamonds), from CNSFR (green asterisks) studied by \cite{2007MNRAS.382..251D} and from central region (dark green triangle) of M33 studied by \cite{2013MNRAS.430..472L} are also included for comparison. Right: Diagnostic diagram of the CALIFA outer region sample (blue circles). Different shades correspond to different bins of distance to the center of the galaxy, as indicated. Point from the IC\,132 \hh region (orange square) of M33 studied by \cite{2013MNRAS.430..472L} is included for comparison. Overplotted as black lines in both diagrams are empirically and theoretically derived separations between LINERs/Seyferts and \hh regions.}
\label{BPT_figure}
\end{figure*}

From the \hh regions catalog published by \cite{2009PhDT........15R} we select
those that fulfil our inner region criteria, specified in Sect. \ref{Inner
  regions sample}, obtaining a total of 79 inner regions, distributed along the
four PINGS galaxies. The specific number of total and inner regions for each
PINGS galaxy are indicated in Table \ref{table_pings_galaxies}. We obtain no
outer regions from the PINGS galaxies, as their spatial linear coverage is
smaller than the one of the CALIFA galaxies and no regions further than 2
$\textit{R}_{eff}$ are extracted.

\begin{table*}
\caption{Physical properties of the PINGS galaxies involved in this work, obtained from the following sources: (i)
Galaxy name. (ii) Redshift, references: NGC\,628, \cite{1993ApJS...88..383L};
NGC\,1058, NGC\,3184, \cite{2005ApJS..160..149S}); NGC\,1637,
\cite{1998AJ....115...62H}. (iii) Morphological type from the R3C catalog,
\cite{1991rc3..book.....D}. (iv) Galaxy inclination angle based on the B25 mag
arcsec$^{-2}$ from R3C. (v) Distances in Mpc, references: NGC\,628,
\cite{2005MNRAS.359..906H}; NGC\,1058, \cite{1996ApJ...466..911E}; NGC\,1637,
\cite{2006ApJS..165..108S}; NGC\,3184, \cite{2002AJ....124.2490L}. (vi)
Effective radius. Derived following the process used by the CALIFA survey,
explained in Sect. \ref{Effective radius}. (vii) and (viii) Magnitudes g and r,
from the Sloan Digital Sky Survey (SDSS). (ix) Absolute B-band magnitude
calculated from the apparent magnitude listed in the R3C catalog and the adopted
distances to the system. (x) Total number of \hh regions extracted in the
galaxy. (xi) Number of inner \hh regions extracted in the galaxy.}  
\label{table_pings_galaxies}    
\centering                                    
\begin{tabular}{l c c c c c c c c }         
\hline\hline                       
\noalign{\smallskip}
              
Galaxy & Redshift & Morph. type & Inclination & Distance & $\textit{R}_{eff}$ & $M_{B}$ & $N_{t}$ & $N_{in}$\\
&&&&(Mpc)&(kpc)&&&\\

\hline

\noalign{\smallskip}

NGC\,628 & 0.0022 & Sc & 24 & 9.3 & 5.51 & -19.9 & 96 & 28\\

NGC\,1058 & 0.0017 & Sc & 21 & 10.6 & 2.31 & -18.3 & 58 & 22\\

NGC\,1637 & 0.0024 & Sc & 36 & 12.0 & 2.83 & -18.9 & 40 & 18\\

NGC\,3184 & 0.0019 & Scd & 21 & 11.1 & 5.88 & -19.9 & 53 & 11\\

\hline
\end{tabular}  
\end{table*}

The distribution of the PINGS inner region sample in the
\oiii~$\lambda$5007/H$\beta$ vs \nii~$\lambda$6583/H$\alpha$ diagram, along with
that from the CALIFA inner regions sample, is shown in Fig. \ref{BPT_figure}. We
can see that the PINGS inner regions follow the same pattern than the CALIFA
inner regions: they have high \nii~$\lambda$6583/H$\alpha$  values, related
to very high oxygen abundances and low percentages of young populations, and
very low \oiii~$\lambda$5007/H$\beta$, due to low excitation values. The PINGS
observations, with higher spatial resolution, allow the detection of inner
regions located closer to their galaxy centers than CALIFA inner
regions. Therefore the location of PINGS inner regions in the BPT diagram show
the continuity of the trend already indicated by CALIFA inner regions.

The comparison with high-resolution circunmnuclear star forming region (CNSFR) observations, that go deeper in the high-metallicity, high-density region around the galactic nucleus, is also a case of great interest. \cite[ hereafter D07]{2007MNRAS.382..251D} studied long-slit observations of 12 CNSFR located in the early-type spiral galaxies NGC\,2903, NGC\,3351 and NGC\,3504. As in the case of PINGS observations, different spatial resolution prevents comparison between properties depending on the region sizes, but does not affect those properties related with the emission-line ratios. Data from these 12 CNSFR are included in the inner regions BPT diagram in Fig. \ref{BPT_figure}, confirming the trend of high oxygen abundances and low excitation values. 

An interesting case of IFS observations of \hh regions located in different environments is the study by \cite[ hereafter LH13]{2013MNRAS.430..472L}, that compares the central region of M33 with IC\,132, a \hh region located at 19 arcmin (4.69 kpc) from the galactic center. These observations were obtained  with the CAHA 3.5-m telescope, using the PMAS instrument in the PPAK mode, as were CALIFA and PINGS observations. Data from the central region and from IC\,132 region are included in the corresponding inner and outer BPT diagrams in Fig. \ref{BPT_figure}. While M33 central region confirms the high-metallicity, low-excitation values indicated by this work and by PINGS and \cite{2007MNRAS.382..251D} data, ID\,132 region expand the trend of the outer regions, with lower metallicity and higher excitation.

\subsubsection{Other diagnostic diagrams}
\label{Disc:Other diagnostic diagrams}

\begin{figure}
\centering
\includegraphics[scale=0.45]{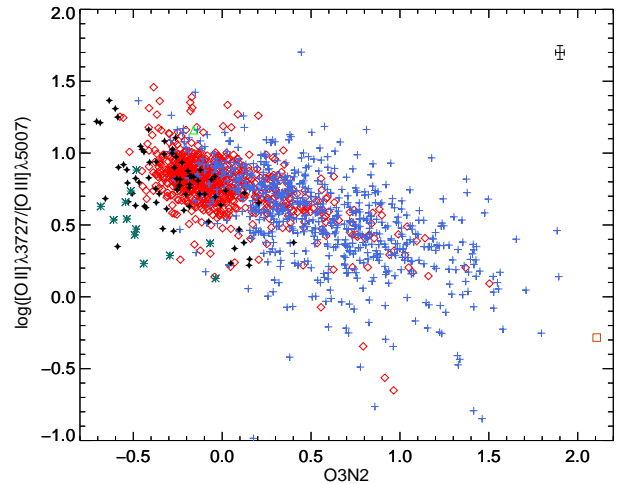}
\caption{Relation between \oii~$\lambda$3727/\oiii~$\lambda$5007 emission-line ratio and the O3N2 index for the CALIFA inner (red diamonds) and outer (blue crosses) regions samples. Data from PINGS inner region sample (black stars), from CNSFR (dark green asterisks) studied by \cite{2007MNRAS.382..251D} and from central region (light green triangle) and IC\,132 \hh region (orange square) of M33 studied by \cite{2013MNRAS.430..472L} are also included for comparison.}
\label{oiioiiivso3n2}
\end{figure}

The study of the relation between several emission-line ratios, that depend on the shape of the ionizing continuum and the physical conditions of the cloud, provide information on physical properties as ages, degree of ionization or abundances. The relation between the \oii~$\lambda$3727/\oiii~$\lambda$5007 emission-line ratio and the O3N2 index, firstly introduced by \cite{1979A&A....78..200A} and defined as

\begin{equation}
{\rm O3N2}=\log \left(\frac{{\rm [O\,III]}\,\lambda5007/H\beta}{{\rm [N\,II]}\,\lambda6584/H\alpha}\right)
,\end{equation}

\noindent
is shown in Fig. \ref{oiioiiivso3n2} for inner and outer CALIFA regions samples. We observe the existence of the same trend between both parameters for both outer and inner regions, although with a difference of one order of magnitude between their values. The O3N2 index, that has an inverse linear relation with the abundance, is smaller for inner regions, due to the weakness of the \oiii~$\lambda$5007 emission line in high metallicity regions. On the other hand outer regions have lower values of the \oii~$\lambda$3727/\oiii~$\lambda$5007 emission-line ratio, denoting a higher degree of ionization than the inner regions sample. As only line ratios are involved and therefore the different spatial resolution has no influence, the PINGS inner regions sample is also included, as well as the CNSFR studied by D07 and M33 central and IC\,132 regions studied by LH13. Their location in the plots confirms and expand the trend followed by the CALIFA inner and outer region samples, as was already seen in the BPT diagrams. 

\begin{figure}
\centering
\includegraphics[scale=0.45]{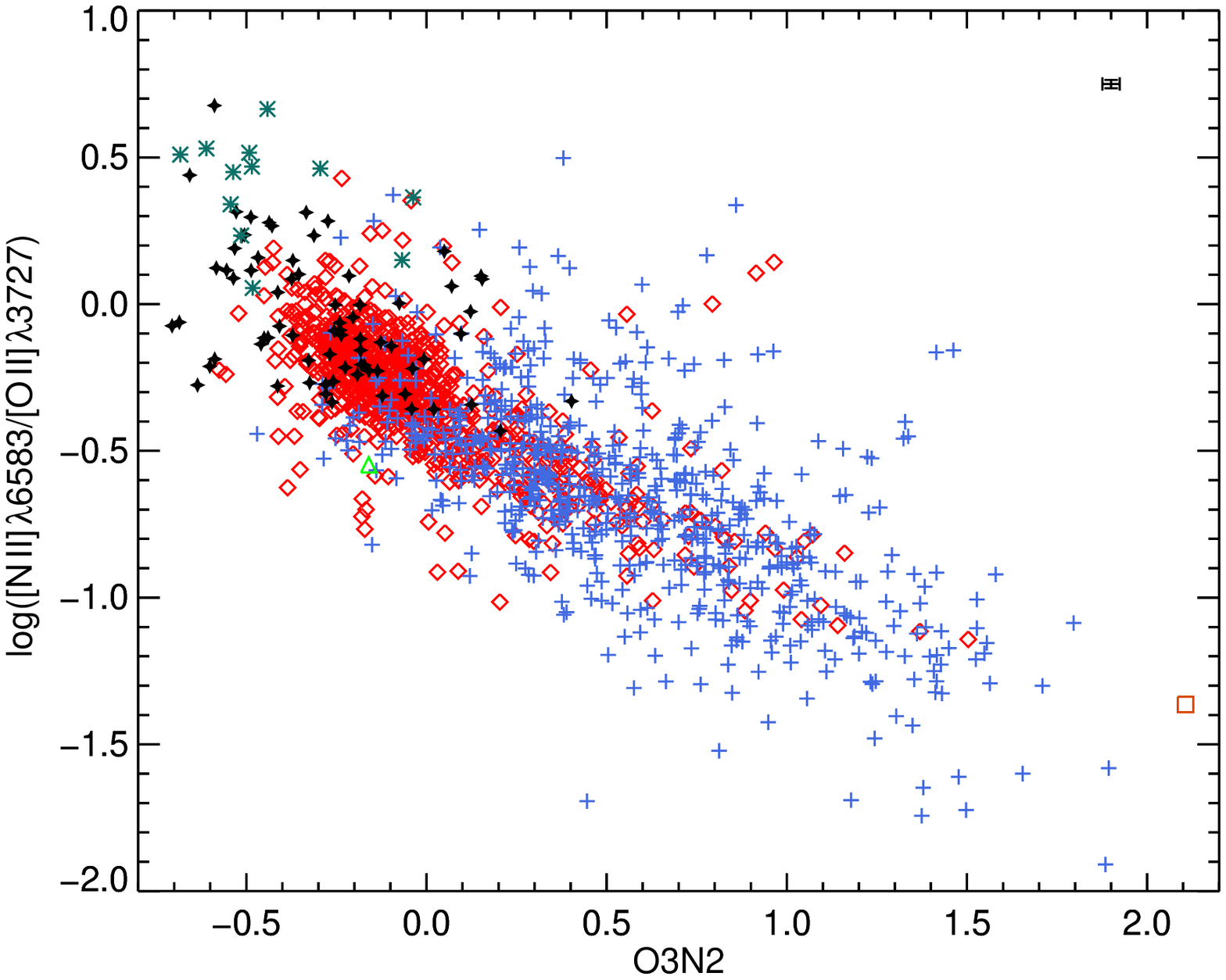}
\caption{Relation between \nii~$\lambda$6583/\oii~$\lambda$3727 emission-line ratio and the O3N2 index for the CALIFA inner (red diamonds) and outer (blue crosses) region samples. Data from PINGS inner region sample (black stars), from CNSFR (dark green asterisks) studied by \cite{2007MNRAS.382..251D} and from central region (light green triangle) and IC\,132 \hh region of M33 (orange square) studied by \cite{2013MNRAS.430..472L} are also included for comparison.}
\label{niioiivso3n2}
\end{figure}

Figure \ref{niioiivso3n2} shows the relation between the \nii~$\lambda$6583/\oii~$\lambda$3727 emission-line ratio and the O3N2 index. The ratio [\ion{N}{ii}]/[\ion{O}{ii}] is a good metallicity indicator: with increasing metallicity the [\ion{O}{ii}] decreases due to the decreasing electron temperature, that prevents the excitation of the O$^{\textit{+}}$ transition, while the [\ion{N}{ii}] does not, therefore causing and increment in [\ion{N}{ii}]/[\ion{O}{ii}] . There is a good correlation between [\ion{N}{ii}]/[\ion{O}{ii}] and the N$^{\textit{+}}$/O$^{\textit{+}}$ ionic abundance ratio, which traces the nitrogen to oxygen abundance ratio \citep{2005MNRAS.361.1063P,2009MNRAS.398..949P}. This correlation was also found by D07 for high metallicity \hh regions and CNSFR. Results observed in Fig. \ref{niioiivso3n2} are in good agreement with in Figs. \ref{BPT_figure} and \ref{oiioiiivso3n2}: inner regions have larger values of [\ion{N}{ii}]/[\ion{O}{ii}], according to their higher metallicity. As in the case of Fig. \ref{oiioiiivso3n2}, the PINGS inner regions sample and the D07 and LH13 data are included in the figure, showing the same trend than the CALIFA inner and outer regions sample.

\begin{figure}
\centering
\includegraphics[scale=0.45]{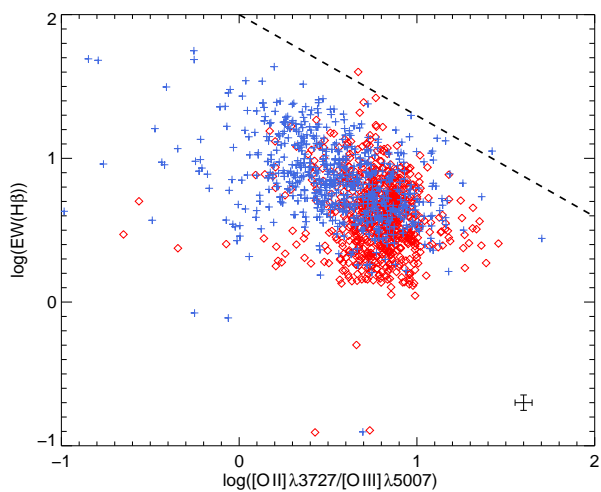}
\caption{Relation between EW(H$\beta$) and \oii~$\lambda$3727/\oiii~$\lambda$5007 emission-line ratio for the inner (red diamonds) and outer (blue crosses) regions samples. Overplotted as a dashed black line is the relation given by \cite{2006MNRAS.365..454H}, explained in the text.}
\label{ewhb_oiioiii}
\end{figure}

The relation between EW(H$\beta$) and \oii~$\lambda$3727/\oiii~$\lambda$5007 emission-line ratio (see Fig. \ref{ewhb_oiioiii}) provides information about the evolution of the star formation processes within a given galaxy. This line ratio is a proxy for the ionization parameter, which in turn is proportional to the quotient of the density of Ly continuum photons to the electron density. The number of hydrogen ionizing photons decreases with the evolution of the ionizing cluster and, other things being equal, lowers the ionization parameter, hence increasing the  \oii~$\lambda$3727/\oiii~$\lambda$5007 line ratio \citep{2006MNRAS.365..454H}. The trend of decreasing EW(H$\beta$), and therefore increasing age for the ionizing population, and increasing \oii~$\lambda$3727/\oiii~$\lambda$5007 line ratio can be seen in Fig. \ref{ewhb_oiioiii} for the observed regions up to EW(H$\beta$) around 3 \AA . Below this value,  which corresponds to ages of the \hh regions of about 10 Myr, we are seeing probably the contribution of an important underlying stellar population which decreases EW(H$\beta$) while keeping the ratio \oii~$\lambda$3727/\oiii~$\lambda$5007 practically constant. The dashed line in the figure marks the envelope of the relation corresponding to the younger \hh region ages and the minimum underlying stellar population contribution (log EW(H$\beta$)=2.00 -0.73 log \oii~$\lambda$3727/\oiii~$\lambda$5007 in \cite{2006MNRAS.365..454H}). Ionizing clusters would beging their evolution from this line downwards, with the starting point depending on their initial mass (higher masses to the left in the plot). Outer regions show larger EW($\beta$) values, related to younger ionizing stellar populations, and smaller \oii~$\lambda$3727/\oiii~$\lambda$5007 values, implying larger ionization parameter values. Since, on average, inner \hh regions have H$\alpha$ luminosities larger than outer ones by about an order of magnitude, this implies that, according to the previous description, the inner regions are ionized by more evolved clusters.

\subsection{Furthest regions}
\label{Furthest regions}

\begin{figure*}
\centering
\includegraphics[scale=0.5]{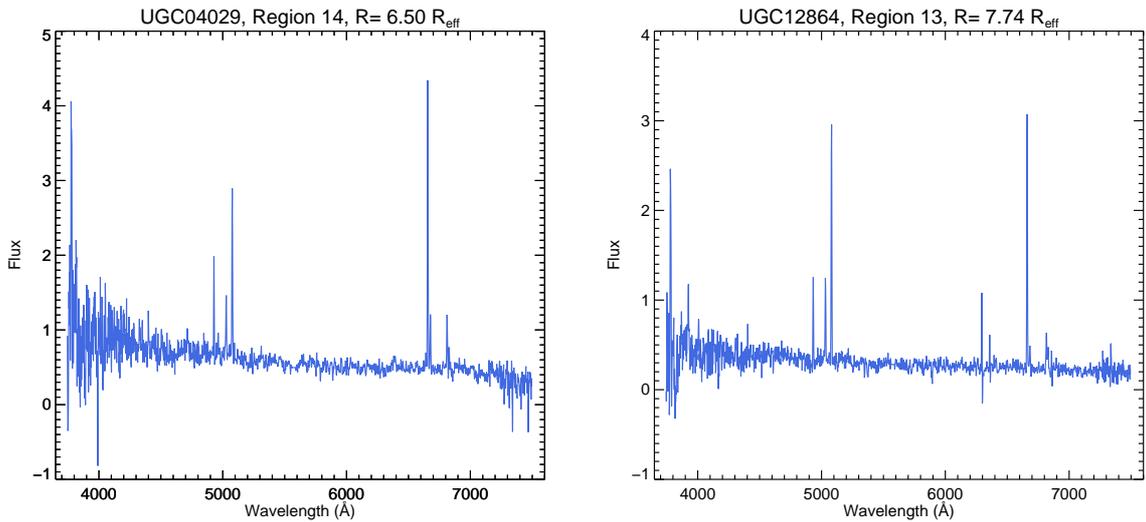}
\caption{Spectra from two of the ten \hh regions located further than 6 $\textit{R}_{eff}$ from their galaxy center, before the subtraction of the stellar continuum. Galaxy name, ID of the region in the galaxy and distance to the center are shown in the titles. Flux is expressed in units of 10$^{-16}$ erg s$^{-1}$ cm$^{-2}$.}
\label{CALIFA_extraouter}
\end{figure*}

The use of IFS techniques allows the detection of \hh regions located much further from the center of the galaxy than it was possible so far \citep[see e.g., ][]{1998AJ....116..673F,1998AJ....116.2805V,2010AJ....139..279W}. In this work 10 of our 671 outer regions are located beyond 6 $\textit{R}_{eff}$, and although five of these ten regions are located close to the projected axis of high inclination galaxies, and therefore have high uncertainties in the determination of their distances to their galactic centers, we consider this group worthy of specific study. Two of these region spectra are included as an example in Fig. \ref{CALIFA_extraouter}, and their most prominent properties are included in Table \ref{table_mostouter_properties}. Mean values of these properties for the whole outer \hh regions sample are also included in the table for comparison. These ten furthest regions were in fact already highlighted in the outer regions BPT diagram in Fig. \ref{BPT_figure}, where it can be observed that they fulfil the general trend followed by the outer regions: they have low \nii~$\lambda$6583/H$\alpha$ emission-line ratio values and high \oiii~$\lambda$5007/H$\beta$ emission-line ratio values, implying low oxygen abundances and high excitations. Other parameters, as EW(H$\alpha$), A$_{V}$ and L(H$\alpha$) also confirm the outer regions trends, as they are in general in good agreement with the outer region average values but slighty above or below, following the corresponding tendency observed in this work for the evolution of the parameter along the galactocentric distance.

\begin{table*}
\caption{Physical properties of outer regions located further than 6 $\textit{R}_{eff}$ from their galaxy center. Emission-line ratios and luminosities are expresed in logarithmic units. Mean values of the outer regions sample are included for comparison.}  
\label{table_mostouter_properties}      
\centering                                    
\begin{tabular}{l c c c c c c c c}       
\hline\hline                      
\noalign{\smallskip}
      
Region ID & Area & \nii~$\lambda$6583/ & \oiii~$\lambda$5007/ & \oii~$\lambda$3727/ & \sii~$\lambda$6717/ & EW(H$\alpha$) & A$_{V}$ & L(H$\alpha$)\\          
& & H$\alpha$ & H$\beta$ & \oiii~$\lambda$5007 & \sii~$\lambda$6731 & & & \\
& (arcsec$^{2}$) & & & & & (\AA) & (mag) & (erg s$^{-1}$) \\

\hline

\noalign{\smallskip}
            
UGC04029-14 & 32 & -0.75 & 0.26 & 0.44 & 0.46 & 60.60 & 0.48 & 38.97\\
UGC04029-19 & 30 & -0.58 & -0.02 & 0.45 & 0.10 & 24.81 & 0.29 & 38.52\\
UGC04029-42 & 26 & -0.67 & 0.03 & 0.36 & 0.27 & 26.49 & 0.13 & 38.23\\
UGC09080-17 & 12 & -1.17 & 0.48 & 0.40 & 0.20 & 43.11 & 0.47 & 37.84\\
UGC12054-21 & 37 & -0.74 & 0.09 & 0.44 & 0.21 & 16.03 & -0.15 & 38.77\\
UGC12054-23 & 32 & -0.80 & 0.02 & 0.64 & 0.51 & 22.86 & 0.77 & 38.81\\
UGC12054-24 & 28 & -0.82 & 0.17 & 0.55 & 0.15 & 12.02 & 0.63 & 38.56\\
UGC12864-13 & 30 & -1.09 & 0.46 & 0.10 & 0.29 & 81.60 & 0.19 & 38.40\\
UGC12864-41 & 27 & -0.85 & 0.18 & 0.31 & 0.10 & 29.44 & 0.05 & 38.48\\
UGC12864-47 & 34 & -0.73 & 0.11 & 0.44 & 0.37 & 23.29 & 0.03 & 38.25\\

& & & & & & & & \\

Outer regions & \multirow{2}{*}{25.6} & \multirow{2}{*}{-0.60} & \multirow{2}{*}{-0.04} & \multirow{2}{*}{0.55} & \multirow{2}{*}{0.17} & \multirow{2}{*}{31.58} & \multirow{2}{*}{0.59} & \multirow{2}{*}{38.85} \\
mean values &  &  &  &  &  &  &  & \\

\hline
\end{tabular}  
\end{table*}

\subsection{Color-magnitude diagrams}
\label{Sec. Colors}

Magnitudes and colors of the outer and inner \hh regions are calculated from their extracted spectra, for a first approach to the spectroscopic properties of the stellar populations. For the calculation of the magnitudes we followed the process indicated in \cite{2009MNRAS.398..451M} and \cite{2013MNRAS.432.2746G}. The filters we considered are the B and V bands from the Johnson's system, and g and r bands from the Sloan SDSS ugriz system, as they are the ones comprised in our data wavelength range. The emission lines considered in the calculation are: \oii~$\lambda$3727, H$\gamma$, H$\beta$, \oiii~$\lambda$4959, \oiii~$\lambda$5007, \hei~$\lambda$5876, \oi~$\lambda$6300, \nii~$\lambda$6548, H$\alpha$, \nii~$\lambda$6583, \sii~$\lambda$6717, and\sii~$\lambda$6731. The transmission curves, the transmission values as a function of wavelength and the transmission values of the broadband filters at the rest wavelength of the selected emission lines employed are those included in the mentioned references. The magnitudes are calculated using the expression: 

\begin{equation}
{\rm m} = -2.5 \log \int_{\lambda_{1}}^{\lambda_{2}} \! L_{\lambda} \, \mathrm{d}\lambda + \sum\limits_{i=1}^{12} \! T_{i} \times L_{i} + C
,\end{equation}

\noindent
where $\lambda_{1}$ and $\lambda_{2}$ are the passband limits in each filter, L$_{\lambda}$ is the stellar SED luminosity, L$_{i}$ is the integrated luminosity in the narrow line for the line \textit{i} and T$_{i}$ the line filter transmission. We assumed that the line width is much narrower that the broadband filter passband. For Johnson's filters, C is the constant for flux calibration in the Vega system. According to the Girardi et al. (2002) prescriptions, Vega is taken as the average of \cite{1997A&AS..125..229L} models for Z = 0.004 and Z = 0.008 at T = 9500K (Vega's data: Z = 0.006, m$_{bol}$ = 0.3319, BC = -0.25 and V = 0.58). C values are 3.27 for B (in this case, where we are calculating B-V) and 2.54 for V. In the SDSS system the constant is always -48.60. Colors are the difference between the selected magnitudes. 

\begin{figure}
\centering
\includegraphics[scale=0.45]{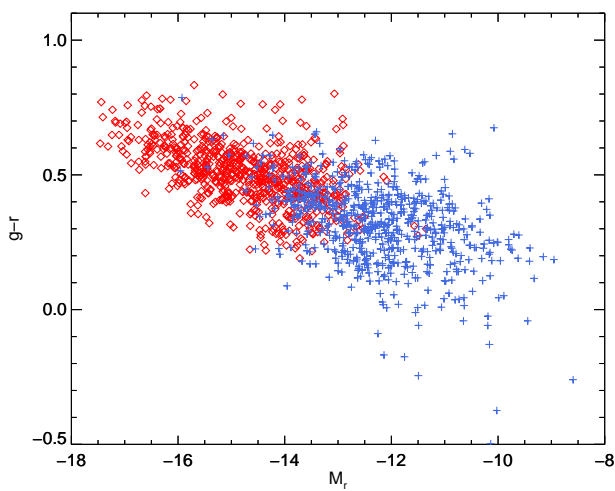}
\caption{Distribution of the inner (red diamonds) and outer (blue crosses) region samples in the g-r vs. \textit{M}$_{\rm r}$ color-magnitude diagram.}
\label{colors}
\end{figure}

In order to remove the contribution of the emission lines, which for the SDSS filter system can imply differences in colors of up to one magnitude for young ages \citep{2013MNRAS.432.2746G}, we calculate the magnitudes masking the mentioned emission lines in the regions spectra. The g-r vs \textit{M}$_{\rm r}$ color-magnitude for both outer and inner samples, calculated for the pure continuum without emission lines, is shown in Fig. \ref{colors}. The color-magnitude diagram of the regions shows that inner regions are redder and have higher luminosities than outer regions, as expected from older regions and with higher metal content, and in good agreement with results obtained in Sects. \ref{Observational and functional parameters} and \ref{Disc:Diagnostic diagrams}.

\subsection{Ionizing and photometric masses}
\label{Ionizing and photometric masses}

The estimation of ionizing and total stellar masses provides more information about the average evolutionary stages for both region samples. Considering the number of ionizing photons for each \hh region calculated in Sect. \ref{Observational and functional parameters}, we estimated the ionizing cluster masses of the regions using the total number of ionizing photons per unit mass provided by the PopStar models \citep{2009MNRAS.398..451M} for a zero-age main sequence with Salpeter initial mass function with lower and upper mass limits of 1 and 100 M$_{\odot}$ and $Z=0.008$. For the inner regions sample we obtain a range of values between $2.43 \times 10^3  - 7.66 \times 10^6 M_{\odot}$, whereas a range of $4.66 \times 10^2  - 7.36 \times 10^5 M_{\odot}$ is obtained for the outer regions sample. In principle, these values are lower limits of the ionizing masses, as we are considering an unevolved stellar population with no photon escape and no dust absorption. The maximum effect of the stellar population evolution can be estimated considering the total number of ionizing photons per unit mass given by the PopStar models with the same IMF and metallicity for a population of 5.2 Myr, as PopStar models consider that clusters older than this age do not produce a visible emission-line ionizing spectrum \citep{2010MNRAS.403.2012M}.  With those conditions the ionizing mass ranges obtained are one order of magnitude larger than those obtained for the zero-age main sequence population. 

\begin{figure}
\centering
\includegraphics[scale=0.45]{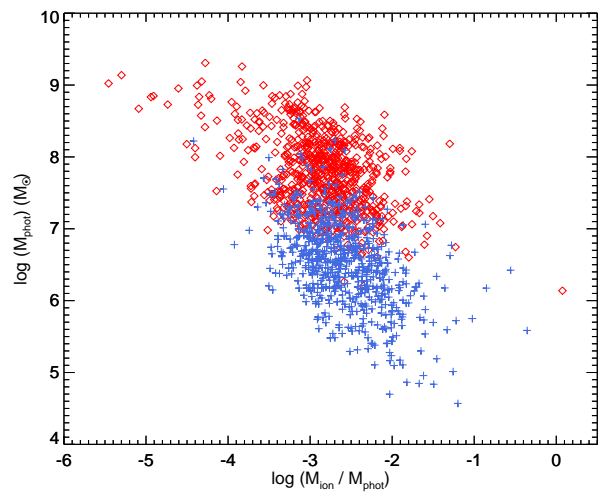}
\caption{Relation between the photometric mass values and the ratio between ionizing and photometric masses for the inner (red diamonds) and outer (blue crosses) regions samples. }
\label{CALIFA_masses}
\end{figure}

One can also estimate photometric masses for the observed regions from the V magnitudes and B-V colors obtained as explained in Sect. \ref{Sec. Colors}, applying the mass-to-light relation described in \cite{2001ApJ...550..212B} for a scaled Salpeter IMF and a formation epoch model with bursts. Figure \ref{CALIFA_masses} shows the relation between the obtained photometric mass values and the ratio between ionizing and photometric masses for both region samples. We observe that inner regions have photometric masses two orders of magnitude bigger than outer regions on average. This is to be expected, due to the higher preponderance of underlying stellar populations from the galaxy bulge in the most internal circumnuclear regions. But again we could also think about a possible selection bias, causing that only the biggest inner regions were detected, but the study of the regions angular areas in Sect. \ref{Observational and functional parameters} already showed that the influence of this bias is small, and therefore there is an intrinsic difference between inner and outer photometric masses. Interestingly enough, the ratios between ionizing masses and photometric masses are similar for inner and outer regions, with the outer regions' ratio values being slightly higher.

\section{Summary and conclusions}
\label{Conclusions}

We have analyzed and compared a sample of 725 inner \hh regions, defined following the criterium by \cite{2015MNRAS.451.3173A}, and a sample of 671 outer \hh regions, located further than 2 $\textit{R}_{eff}$ from their corresponding galactic center. The \hh regions were detected and extracted applying the {\sc HIIexplorer} procedure \citep{2012A&A...546A...2S,2012ApJ...756L..31R} to the observations of a sample of 263 isolated spiral galaxies, part of the CALIFA survey \citep{2012A&A...538A...8S}. 

Different trends and values of the main physical properties of \hh regions are
observed in the comparison between inner and outer regions samples. Inner
regions show lower hydrogen line equivalent width
values, higher extinction and higher luminosities and number of ionizing
photons, as well as larger values of \nii~$\lambda$6583/H$\alpha$ and
\oii~$\lambda$3727/\oiii~$\lambda$5007 line ratios, related to higher oxygen abundances and
smaller ionization parameters respectively. According to these facts we conclude
that inner regions have more evolved stellar populations and are in a later
evolution state with respect to the outer regions. The distribution of both
region samples across several diagnostic diagrams confirm this conclusion. 

We have calculated magnitudes and colors from the regions extracted spectra, observing that inner regions are redder and have higher luminosities, as expected. We have also estimated the photometric stellar masses and the ionizing stellar masses of the regions, obtaining higher masses for the inner regions and slightly higher M$_{ion}$/M$_{phot}$  values for the outer regions. 

This characterization of observational properties of two homogeneous and
coherent inner and outer regions samples confirm and expand previous results
about intrinsic differences depending on the location of the regions and the
influence of the environment, related to different evolution stages and
therefore providing information about the formation and evolution processes of
the galaxies. These different aspects will be further explored in the second paper of this series 
by combining both stellar population and photoionization models.

\begin{acknowledgements} We acknowledge financial support for the ESTALLIDOS collaboration by the Spanish Ministerio de Econom\'{i}a y Competitividad (MINECO) under grant AYA2013-47742-C4-3-P. We acknowledge financial support from the Marie Curie FP7-PEOPLE-2013-IRSES scheme, under the SELGIFS collaboration (Study of Emission-Line Galaxies with Integral-Field Spectroscopy). M.~R.~B. acknowledges financial support by the Spanish Ministerio de Econom\'{i}a y Competitividad under the FPI fellowships program. 

\end{acknowledgements}

{}


\begin{thebibliography}{}
\bibliographystyle{aa}


\bibitem[Alloin et al.(1979)]{1979A&A....78..200A} Alloin, D., Collin-Souffrin, S., Joly, M., \& Vigroux, L.\ 1979, \aap, 78, 200 

\bibitem[{\'A}lvarez-{\'A}lvarez et al.(2015)]{2015MNRAS.451.3173A} {\'A}lvarez-{\'A}lvarez, M., D{\'{\i}}az, A.~I., Terlevich, E., \& Terlevich, R.\ 2015, \mnras, 451, 3173 


\bibitem[Baldwin et al.(1981)]{1981PASP...93....5B} Baldwin, J.~A., Phillips, M.~M., \& Terlevich, R.\ 1981, \pasp, 93, 5 

\bibitem[Baldwin et al.(1991)]{1991ApJ...374..580B} Baldwin, J.~A., Ferland, G.~J., Martin, P.~G., et al.\ 1991, \apj, 374, 580 

\bibitem[Bell \& de Jong(2001)]{2001ApJ...550..212B} Bell, E.~F., \& de Jong, R.~S.\ 2001, \apj, 550, 212 

\bibitem[Berg et al.(2015)]{2015ApJ...806...16B} 
Berg, D.~A., Skillman, E.~D., Croxall, K.~V., et al.\ 2015, \apj, 806, 16

\bibitem[Bresolin \& Kennicutt(2015)]{2015MNRAS.454.3664B} Bresolin, F., \& Kennicutt, R.~C.\ 2015, \mnras, 454, 3664 

\bibitem[Bresolin et al.(2004)]{2004ApJ...615..228B} Bresolin, F., Garnett, D.~R., \& Kennicutt, R.~C., Jr.\ 2004, \apj, 615, 228 



\bibitem[Cardelli et al.(1989)]{1989ApJ...345..245C} Cardelli, J.~A., Clayton, G.~C., \& Mathis, J.~S.\ 1989, \apj, 345, 245 


\bibitem[Cid Fernandes et al.(2010)]{2010MNRAS.403.1036C} Cid Fernandes, R., Stasi{\'n}ska, G., Schlickmann, M.~S., et al.\ 2010, \mnras, 403, 1036 

\bibitem[Cid Fernandes et al.(2013)]{2013A&A...557A..86C} Cid Fernandes, R., P{\'e}rez, E., Garc{\'{\i}}a Benito, R., et al.\ 2013, \aap, 557, A86 

\bibitem[Cid Fernandes et al.(2014)]{2014A&A...561A.130C} Cid Fernandes, R., Gonz{\'a}lez Delgado, R.~M., Garc{\'{\i}}a Benito, R., et al.\ 2014, \aap, 561, A130 

\bibitem[Contini et al.(1998)]{1998A&AS..130..285C} Contini, T., Considere, S., \& Davoust, E.\ 1998, \aaps, 130, 285 



\bibitem[de Vaucouleurs et al.(1991)]{1991rc3..book.....D} de Vaucouleurs, G., de Vaucouleurs, A., Corwin, H.~G., Jr., et al.\ 1991, Third Reference Catalogue of Bright Galaxies.

\bibitem[D{\'{\i}}az(1989)]{1989epg..conf..377D} D{\'{\i}}az, {\'A}.~I.\ 1989, Evolutionary Phenomena in Galaxies, 377 

\bibitem[D{\'{\i}}az(1998)]{1998Ap&SS.263..143D} D{\'{\i}}az, {\'A}.~I.\ 1998, \apss, 263, 143 

\bibitem[D{\'{\i}}az et al.(2000)]{2000MNRAS.318..462D} D{\'{\i}}az, A.~I., Castellanos, M., Terlevich, E., \& Luisa Garc{\'{\i}}a-Vargas, M.\ 2000, \mnras, 318, 462 

\bibitem[D{\'{\i}}az et al.(2007)]{2007MNRAS.382..251D} D{\'{\i}}az, {\'A}.~I., Terlevich, E., Castellanos, M., \& H{\"a}gele, G.~F.\ 2007, \mnras, 382, 251 



\bibitem[Eastman et al.(1996)]{1996ApJ...466..911E} Eastman, R.~G., Schmidt, B.~P., \& Kirshner, R.\ 1996, \apj, 466, 911 



\bibitem[Falc{\'o}n-Barroso et al.(2011)]{2011A&A...532A..95F} Falc{\'o}n-Barroso, J., S{\'a}nchez-Bl{\'a}zquez, P., Vazdekis, A., et al.\ 2011, \aap, 532, A95 

\bibitem[Ferguson et al.(1998)]{1998AJ....116..673F} Ferguson, A.~M.~N., Gallagher, J.~S., \& Wyse, R.~F.~G.\ 1998, \aj, 116, 673 

\bibitem[Freeman(1970)]{1970ApJ...160..811F} Freeman, K.~C.\ 1970, \apj, 160, 811 



\bibitem[Garc{\'{\i}}a-Benito et al.(2015)]{2015A&A...576A.135G} Garc{\'{\i}}a-Benito, R., Zibetti, S., S{\'a}nchez, S.~F., et al.\ 2015, \aap, 576, A135 

\bibitem[Garc{\'{\i}}a-Vargas et al.(2013)]{2013MNRAS.432.2746G} Garc{\'{\i}}a-Vargas, M.~L., Moll{\'a}, M., \& Mart{\'{\i}}n-Manj{\'o}n, M.~L.\ 2013, \mnras, 432, 2746 

\bibitem[Girardi et al.(2002)]{2002A&A...391..195G} Girardi, L., Bertelli, G., Bressan, A., et al.\ 2002, \aap, 391, 195 

\bibitem[Gonz{\'a}lez Delgado \& P{\'e}rez(1997)]{1997ApJS..108..199G} Gonz{\'a}lez Delgado, R.~M., \& P{\'e}rez, E.\ 1997, \apjs, 108, 199 

\bibitem[Gonzalez-Delgado et al.(1995)]{1995ApJ...439..604G} 
Gonzalez-Delgado, R.~M., Perez, E., Diaz, A.~I., et al.\ 1995, \apj, 439, 
604 




\bibitem[Haynes et al.(1998)]{1998AJ....115...62H} Haynes, M.~P., van Zee, L., Hogg, D.~E., Roberts, M.~S., \& Maddalena, R.~J.\ 1998, \aj, 115, 62 

\bibitem[Heidmann et al.(1972)]{1972MmRAS..75...85H} Heidmann, J., Heidmann, N., \& de Vaucouleurs, G.\ 1972, \memras, 75, 85 

\bibitem[Hendry et al.(2005)]{2005MNRAS.359..906H} Hendry, M.~A., Smartt, S.~J., Maund, J.~R., et al.\ 2005, \mnras, 359, 906 

\bibitem[Henry(1993)]{1993MNRAS.261..306H} Henry, R.~B.~C.\ 1993, \mnras, 261, 306 

\bibitem[Holmberg(1958)]{1958MeLuS.136....1H} Holmberg, E.\ 1958, Meddelanden fran Lunds Astronomiska Observatorium Serie II, 136, 1 

\bibitem[Hoyos \& D{\'{\i}}az(2006)]{2006MNRAS.365..454H} Hoyos, C., \& D{\'{\i}}az, A.~I.\ 2006, \mnras, 365, 454 

\bibitem[Husemann et al.(2013)]{2013A&A...549A..87H} Husemann, B., Jahnke, K., S{\'a}nchez, S.~F., et al.\ 2013, \aap, 549, A87 



\bibitem[Kauffmann et al.(2003)]{2003MNRAS.346.1055K} Kauffmann, G., 
Heckman, T.~M., Tremonti, C., et al.\ 2003, \mnras, 346, 1055 

\bibitem[Kehrig et al.(2012)]{2012A&A...540A..11K} Kehrig, C., Monreal-Ibero, A., Papaderos, P., et al.\ 2012, \aap, 540, A11 

\bibitem[Kelz \& Roth(2006)]{2006NewAR..50..355K} Kelz, A., \& Roth, M.~M.\ 2006, \nar, 50, 355 

\bibitem[Kelz et al.(2006)]{2006PASP..118..129K} Kelz, A., Verheijen, M.~A.~W., Roth, M.~M., et al.\ 2006, \pasp, 118, 129 

\bibitem[Kennicutt et al.(1989)]{1989AJ.....97.1022K} Kennicutt, R.~C., Jr., Keel, W.~C., \& Blaha, C.~A.\ 1989, \aj, 97, 1022 

\bibitem[Kewley et al.(2001)]{2001ApJ...556..121K} Kewley, L.~J., Dopita, 
M.~A., Sutherland, R.~S., Heisler, C.~A., 
\& Trevena, J.\ 2001, \apj, 556, 121 


\bibitem[Lejeune et al.(1997)]{1997A&AS..125..229L} Lejeune, T., Cuisinier, F., \& Buser, R.\ 1997, \aaps, 125,  

\bibitem[Leonard et al.(2002)]{2002AJ....124.2490L} Leonard, D.~C., Filippenko, A.~V., Li, W., et al.\ 2002, \aj, 124, 2490 

\bibitem[Lopez et al.(2011)]{2011ApJ...731...91L} Lopez, L.~A., Krumholz, M.~R., Bolatto, A.~D., Prochaska, J.~X., \& Ramirez-Ruiz, E.\ 2011, \apj, 731, 91 

\bibitem[L{\'o}pez-Hern{\'a}ndez et al.(2013)]{2013MNRAS.430..472L} L{\'o}pez-Hern{\'a}ndez, J., Terlevich, E., Terlevich, R., et al.\ 2013, \mnras, 430, 472 

\bibitem[Lu et al.(1993)]{1993ApJS...88..383L} Lu, N.~Y., Hoffman, G.~L., Groff, T., Roos, T., \& Lamphier, C.\ 1993, \apjs, 88, 383 



\bibitem[Makarov et al.(2014)]{2014A&A...570A..13M} Makarov, D., Prugniel, P., Terekhova, N., Courtois, H., \& Vauglin, I.\ 2014, \aap, 570, A13 

\bibitem[Marino et al.(2012)]{2012ApJ...754...61M} Marino, R.~A., Gil de Paz, A., Castillo-Morales, A., et al.\ 2012, \apj, 754, 61 

\bibitem[Marino et al.(2016)]{2016A&A...585A..47M} Marino, R.~A., Gil de Paz, A., S{\'a}nchez, S.~F., et al.\ 2016, \aap, 585, A47 

\bibitem[M{\'a}rmol-Queralt{\'o} et al.(2011)]{2011A&A...534A...8M} M{\'a}rmol-Queralt{\'o}, E., S{\'a}nchez, S.~F., Marino, R.~A., et al.\ 2011, \aap, 534, A8 

\bibitem[Mart{\'{\i}}n-Manj{\'o}n et al.(2010)]{2010MNRAS.403.2012M} Mart{\'{\i}}n-Manj{\'o}n, M.~L., Garc{\'{\i}}a-Vargas, M.~L., Moll{\'a}, M., \& D{\'{\i}}az, A.~I.\ 2010, \mnras, 403, 2012 

\bibitem[Mas-Hesse \& Kunth(1991)]{1991A&AS...88..399M} Mas-Hesse, J.~M., \& Kunth, D.\ 1991, \aaps, 88, 399 

\bibitem[Mast et al.(2014)]{2014A&A...561A.129M} Mast, D., Rosales-Ortega, F.~F., S{\'a}nchez, S.~F., et al.\ 2014, \aap, 561, A129 

\bibitem[Mathis(1986)]{1986PASP...98..995M} Mathis, J.~S.\ 1986, \pasp, 98, 995 

\bibitem[Miralles-Caballero et al.(2014)]{2014MNRAS.440.2265M} 
Miralles-Caballero, D., D{\'{\i}}az, A.~I., Rosales-Ortega, F.~F., 
P{\'e}rez-Montero, E., \& S{\'a}nchez, S.~F.\ 2014, \mnras, 440, 2265

\bibitem[Moll{\'a} et al.(2009)]{2009MNRAS.398..451M} Moll{\'a}, M., Garc{\'{\i}}a-Vargas, M.~L., \& Bressan, A.\ 2009, \mnras, 398, 451 



\bibitem[Oey \& Kennicutt(1993)]{1993ApJ...411..137O} Oey, M.~S., \& Kennicutt, R.~C., Jr.\ 1993, \apj, 411, 137 

\bibitem[Oey et al.(2003)]{2003AJ....126.2317O} Oey, M.~S., Parker, J.~S., Mikles, V.~J., \& Zhang, X.\ 2003, \aj, 126, 2317 

\bibitem[Opik(1923)]{1923Obs....46...51O} Opik, E.\ 1923, The Observatory, 46, 51 

\bibitem[Osterbrock 
\& Ferland(2006)]{2006agna.book.....O} Osterbrock, D.~E., \& Ferland, G.~J.\ 2006, Astrophysics of gaseous nebulae and active galactic nuclei, 2nd.~ed.~by D.E.~Osterbrock and G.J.~Ferland.~Sausalito, CA: University Science Books, 2006



\bibitem[P{\'e}rez-Montero \& Contini(2009)]{2009MNRAS.398..949P} P{\'e}rez-Montero, E., \& Contini, T.\ 2009, \mnras, 398, 949 

\bibitem[P{\'e}rez-Montero \& D{\'{\i}}az(2005)]{2005MNRAS.361.1063P} P{\'e}rez-Montero, E., \& D{\'{\i}}az, A.~I.\ 2005, \mnras, 361, 1063 

\bibitem[Perinotto(1983)]{1983ASIC..110..205P} Perinotto, M.\ 1983, NATO Advanced Science Institutes (ASI) Series C, 110, 205 



\bibitem[Rosales-Ortega(2009)]{2009PhDT........15R} Rosales-Ortega, F.~F.\ 
2009, Ph.D.~Thesis

\bibitem[Rosales-Ortega et al.(2010)]{2010MNRAS.405..735R} Rosales-Ortega, 
F.~F., Kennicutt, R.~C., S{\'a}nchez, S.~F., et al.\ 2010, \mnras, 405, 735 

\bibitem[Rosales-Ortega et al.(2011)]{2011MNRAS.415.2439R} Rosales-Ortega, F.~F., D{\'{\i}}az, A.~I., Kennicutt, R.~C., \& S{\'a}nchez, S.~F.\ 2011, \mnras, 415, 2439 

\bibitem[Rosales-Ortega et al.(2012)]{2012ApJ...756L..31R} Rosales-Ortega, F.~F., S{\'a}nchez, S.~F., Iglesias-P{\'a}ramo, J., et al.\ 2012, \apjl, 756, L31 

\bibitem[Rosolowsky \& Simon (2008)]{2008ApJ...675.1213R}
Rosolowsky, E., Simon, J.~ D.,\ 2008, \apj, 675, 1213

\bibitem[Roth et al.(2005)]{2005PASP..117..620R} Roth, M.~M., Kelz, A., Fechner, T., et al.\ 2005, \pasp, 117, 620 


\bibitem[Saha et al.(2006)]{2006ApJS..165..108S} Saha, A., Thim, F., Tammann, G.~A., Reindl, B., \& Sandage, A.\ 2006, \apjs, 165, 108 

\bibitem[S{\'a}nchez et al.(2006)]{2006AN....327..167S} S{\'a}nchez, S.~F., Garc{\'{\i}}a-Lorenzo, B., Jahnke, K., et al.\ 2006, Astronomische Nachrichten, 327, 167 

\bibitem[S{\'a}nchez et al.(2011)]{2011MNRAS.410..313S} S{\'a}nchez, S.~F., Rosales-Ortega, F.~F., Kennicutt, R.~C., et al.\ 2011, \mnras, 410, 313 

\bibitem[S{\'a}nchez et al.(2012)]{2012A&A...538A...8S} S{\'a}nchez, S.~F., Kennicutt, R.~C., Gil de Paz, A., et al.\ 2012, \aap, 538, A8 

\bibitem[S{\'a}nchez et al.(2012)]{2012A&A...546A...2S} S{\'a}nchez, S.~F., Rosales-Ortega, F.~F., Marino, R.~A., et al.\ 2012, \aap, 546, A2 

\bibitem[S{\'a}nchez et al.(2014)]{2014A&A...563A..49S} S{\'a}nchez, S.~F., Rosales-Ortega, F.~F., Iglesias-P{\'a}ramo, J., et al.\ 2014, \aap, 563, A49 

\bibitem[S{\'a}nchez et al.(2015)]{2015A&A...574A..47S} S{\'a}nchez, S.~F., P{\'e}rez, E., Rosales-Ortega, F.~F., et al.\ 2015, \aap, 574, A47 

\bibitem[S{\'a}nchez et al.(2016)]{2016A&A...594A..36S} S{\'a}nchez, S.~F., Garc{\'{\i}}a-Benito, R., Zibetti, S., et al.\ 2016, \aap, 594, A36 

\bibitem[S{\'a}nchez et al.(2016)]{2016RMxAA..52...21S} S{\'a}nchez, S.~F., P{\'e}rez, E., S{\'a}nchez-Bl{\'a}zquez, P., et al.\ 2016, \rmxaa, 52, 21 

\bibitem[S{\'a}nchez et al.(2016)]{2016RMxAA..52..171S} S{\'a}nchez, S.~F., P{\'e}rez, E., S{\'a}nchez-Bl{\'a}zquez, P., et al.\ 2016, \rmxaa, 52, 171 

\bibitem[S{\'a}nchez-Menguiano et al.(2016)]{2016ApJ...830L..40S} S{\'a}nchez-Menguiano, L., S{\'a}nchez, S.~F., Kawata, D., et al.\ 2016, \apjl, 830, L40 

\bibitem[Schawinski et al.(2014)]{2014MNRAS.440..889S} Schawinski, K., Urry, C.~M., Simmons, B.~D., et al.\ 2014, \mnras, 440, 889 

\bibitem[Springob et al.(2005)]{2005ApJS..160..149S} Springob, C.~M., Haynes, M.~P., Giovanelli, R., \& Kent, B.~R.\ 2005, \apjs, 160, 149 



\bibitem[van Zee et al.(1998)]{1998AJ....116.2805V} van Zee, L., Salzer, J.~J., Haynes, M.~P., O'Donoghue, A.~A., \& Balonek, T.~J.\ 1998, \aj, 116, 2805 

\bibitem[Vazdekis et al.(2010)]{2010MNRAS.404.1639V} Vazdekis, A., S{\'a}nchez-Bl{\'a}zquez, P., Falc{\'o}n-Barroso, J., et al.\ 2010, \mnras, 404, 1639 

\bibitem[Verheijen et al.(2004)]{2004AN....325..151V} Verheijen, M.~A.~W., Bershady, M.~A., Andersen, D.~R., et al.\ 2004, Astronomische Nachrichten, 325, 151 

\bibitem[V{\'e}ron-Cetty \& V{\'e}ron(2006)]{2006A&A...455..773V} V{\'e}ron-Cetty, M.-P., \& V{\'e}ron, P.\ 2006, \aap, 455, 773 


\bibitem[Walcher et al.(2014)]{2014A&A...569A...1W} Walcher, C.~J., Wisotzki, L., Bekerait{\'e}, S., et al.\ 2014, \aap, 569, A1 

\bibitem[Werk et al.(2010)]{2010AJ....139..279W} Werk, J.~K., Putman, M.~E., Meurer, G.~R., et al.\ 2010, \aj, 139, 279-295 

\bibitem[Willett et al.(2013)]{2013MNRAS.435.2835W} Willett, K.~W., Lintott, C.~J., Bamford, S.~P., et al.\ 2013, \mnras, 435, 2835 


\bibitem[York et al.(2000)]{2000AJ....120.1579Y} York, D.~G., Adelman, J., Anderson, J.~E., Jr., et al.\ 2000, \aj, 120, 1579 


\end{thebibliography}
\end{document}